\documentclass[cits,hyper]{JINST}
\usepackage{isotope}
\usepackage{multirow}

\title{Measurement of the quenching factor of Na recoils in NaI(Tl)}

\author{H.~Chagani\thanks{Corresponding author}, P.~Majewski,
  E.~J.~Daw, V.~A.~Kudryavtsev \& N.~J.~C.~Spooner \\
  Department of Physics and Astronomy, University of Sheffield, Hicks Building,
  Hounsfield Road, Sheffield S3 7RH, United Kingdom \\
  E-mail: \email{hassan.chagani@ijs.si}}

\abstract{Measurements of the quenching factor for sodium recoils in a 5~cm
  diameter NaI(Tl) crystal at room temperature have been made at a dedicated
  neutron facility at the University of Sheffield. The crystal has been
  exposed to 2.45~MeV mono-energetic neutrons generated by a Sodern GENIE~16
  neutron generator, yielding nuclear recoils of energies between 10~and
  100~keVnr. A cylindrical BC501A detector has been used to tag neutrons that
  scatter off sodium nuclei in the crystal. Cuts on pulse shape and time of
  flight have been performed on pulses recorded by an Acqiris DC265 digitiser
  with a 2~ns sampling time. Measured quenching factors of Na nuclei range
  from 19\% to 26\% in good agreement with other experiments, and a value of
  $25.2 \pm 6.4$\% has been determined for 10~keV sodium recoils. From pulse
  shape analysis, the mean times of pulses from electron and nuclear recoils
  have been compared down to 2~keVee. The experimental results are compared to
  those predicted by Lindhard theory, simulated by the SRIM Monte Carlo code,
  and a preliminary curve calculated by Prof.~Akira Hitachi.}


\begin{document}

\section{Introduction}

Astronomical observations, such as galactic rotation curves~\cite{Sofue01} and
gravitational lensing~\cite{Cabanac05}, combined with measurements of the
temperature fluctuations in the Cosmic Microwave Background~\cite{Spergel07}
and abundances of light nuclei~\cite{Burles01}, point to the striking
conclusion that the majority of matter in the Universe does not consist of the
stars, planets and gas that are visible in the images from telescopes. A
possible solution to this is the presence of a more elusive particle population
of `Dark Matter', that contributes to most of the mass of galaxies. Earth-based
detectors for dark matter particles passing through the Earth typically
utilise large masses of ultra-radiopure target materials, in what is referred
to as the direct method. Of the many possible candidates, the Weakly
Interacting Massive Particle (WIMP) has the most direct search experiments
dedicated to its discovery. Direct searches for WIMPs detect the elastic
recoil of an incident WIMP off a target nucleus. Such an interaction deposits
a recoil energy $E_{R}$ in the detector. A variety of approaches to the direct
detection of dark matter are adopted by various international collaborations.
A recent comprehensive review is given by~\cite{Spooner07}.

Inorganic crystal scintillators are popular choices as target materials for
direct dark matter search experiments. The high light yield and pulse shape
differences between nuclear and electron recoils explain why thallium activated
sodium iodide (NaI(Tl)) crystals are the oldest scintillators used in such
experiments~\cite{Quenby95}. They still remain one of the best detectors at
determining spin-dependent WIMP-nucleon limits, and the ANAIS~\cite{Amare06},
DAMA/NaI~\cite{Bernabei98} and ELEGANT-V~\cite{Fushimi99} direct search
experiments utilise them. The DAMA/NaI experiment is the only one that has
claimed to witness the annual modulation of a WIMP
signal~\cite{Bernabei98}~\cite{Bernabei00}, and until recently,
NAIAD~\cite{Ahmed03} held the best spin-dependent limit on WIMP-proton
interactions~\cite{Alner05}. DAMA/LIBRA~\cite{Bernabei06} is a next generation
NaI(Tl)-based detector currently taking data at Gran Sasso. Hence, NaI(Tl)
remains an important detector material in non-baryonic dark matter searches.

Energy scale calibration is performed by exposing the detector to radiation
from a gamma-ray emitting radioisotope. Unlike neutrons and WIMPs, detectable
energy from gamma-rays is a result of collisions with target electrons rather
than nuclei. The energy deposited by nuclear recoils is less than that for
electron recoils of the same $E_{R}$, which is known as ionisation
quenching~\cite{Birks64}~\cite{Voltz66}. In other words,
$E_{\mathrm{vis}} = QE_{R}$, where $E_{\mathrm{vis}}$ is the visible energy,
and $Q$ is the measurable quantity showing the degree of quenching for nuclear
recoils with respect to electron interactions, also known as the quenching
factor. When calculating the WIMP-nucleon differential cross-section to derive
a limit as outlined in~\cite{Lewin96}, this effect can be corrected for by
multiplying the detected energy by the reciprocal of the quenching factor. It
is necessary to determine the quenching factor for each scintillating dark
matter target independently. Additionally, the scintillation efficiency
changes depending on the recoil energy, and combined with form factor
corrections to the WIMP-nucleon differential cross-section that favour low
energy recoils~\cite{Lewin96}, it is important to conduct measurements at
energies relevant to dark matter searches (below 50~keV).

Quenching factors of Na recoils in NaI(Tl) have been measured
by~\cite{Spooner94}~\cite{Tovey98}~\cite{Gerbier99}~\cite{Simon03} to a
minimum recoil energy of 15~keV. The experiment described here has probed the
quenching factor to a lower recoil energy of 10~keV, and has achieved the
highest accuracy above 20~keV. Between the energy range 10~to 100~keV, it
provides the most detailed measurement of the quenching factor to date.

\section{Theoretical overview}\label{theory_sec}
After a nuclear interaction, a recoiling nucleus will lose energy as it moves
through a target material through collisions with electrons (hereafter called
electronic energy loss, resp. electronic energy loss mechanism) and other
nuclei. As most detectors, including scintillators, are sensitive to
electronic energy loss only, the quenching factor can be calculated through an
understanding of these mechanisms. In other words, scintillation light can be
understood to be the result of the electronic energy loss mechanisms, while
non-radiative transfers, such as heat, are due to collisions with other
nuclei. The Lindhard theory~\cite{Lindhard63_1}~\cite{Lindhard63_2} attempts
to quantify these interactions from first principles, and the points relevant
to the theoretical determination of the quenching factor are briefly outlined
here.

The energy loss mechanisms through the electronic and nuclear channels can be
understood as the electronic and nuclear stopping powers respectively. These
can be defined by rescaling the range $R$ and energy $E_R$ of a recoiling
nucleus to the respective non-dimensional variables $\rho$ and
$\epsilon$~\cite{Lindhard63_1}. In such a way, the nuclear energy loss
$\left(\frac{d\epsilon}{d\rho}\right)_{\mathrm{n}}$ can be defined as a
universal function $f(\epsilon)$ that can be calculated numerically.

When the penetrating particle and the atoms of the medium are the same,
$\epsilon$ becomes:

\begin{equation}\label{epsilon_eq}
\epsilon = \frac{11.5}{Z^{\frac{7}{3}}} E_{R}
\end{equation}
where $Z$ is the atomic number of the target nuclei and $E_{R}$ is the
deposited energy in keV.

The electronic energy loss is defined by
$\left(\frac{d\epsilon}{d\rho}\right)_{\mathrm{e}} = \kappa\sqrt{\epsilon}$.
If the penetrating particle is identical to the atoms of the medium, the
constant $\kappa$ is given by:

\begin{equation}\label{kappa_eq}
\kappa = \frac{0.133 Z^{\frac{1}{2}}}{\sqrt{A}} \xi_{\mathrm{e}}
\end{equation}
where $A$ is the mass number of the target nuclei and
$\xi_{\mathrm{e}}\approx Z^{\frac{1}{6}}$ from~\cite{Lindhard63_1}.

Assuming that the electronic and nuclear collisions are uncorrelated, the total
energy given to electrons and that given to atoms can be expressed as the two
separate quantities $\eta$ and $\nu$ respectively. The non-dimensional
variable $\epsilon$ can now be written in terms of these:

\begin{equation}\label{ERlind_eq}
\epsilon = \bar{\eta} + \bar{\nu}
\end{equation}

For large $\epsilon$, the mean energy given to atoms of the medium $\bar{\nu}$
is inversely proportional to $\kappa$. However, this does not hold when
$\epsilon < 1$, in which case $\bar{\nu}\approx\epsilon$. A single formula that
combines these results is~\cite{Lindhard63_2}:

\begin{equation}\label{nubar_eq}
\bar{\nu} = \frac{\epsilon}{1 + \kappa g(\epsilon)}
\end{equation}
where the function $g(\epsilon)$ is well fitted by~\cite{Lewin96}:

\begin{equation}\label{ge_eq}
g(\epsilon) = 3\epsilon^{0.15} + 0.7\epsilon^{0.6} + \epsilon
\end{equation}

From Eq.~(\ref{ERlind_eq}), the mean energy given to electrons $\bar{\eta}$ in
terms of $\epsilon$ can be written as $\bar{\eta} = \epsilon - \bar{\nu}$.
Therefore, an expression for the quenching factor can be obtained using its
definition as given previously, by dividing $\bar{\eta}$ by $\epsilon$ through
a combination of Eq.~(\ref{ERlind_eq}) and Eq.~(\ref{nubar_eq}):

\begin{equation}\label{QFlind_eq}
\frac{\bar{\eta}}{\epsilon} = \frac{\epsilon - \bar{\nu}}{\epsilon} =
\frac{\epsilon(1 + \kappa g(\epsilon)) - \epsilon}
{\epsilon(1 + \kappa g(\epsilon))} =
\frac{\kappa g(\epsilon)}{1 + \kappa g(\epsilon)}
\end{equation}

\begin{figure}
  \begin{center}
    \includegraphics[scale=0.7]{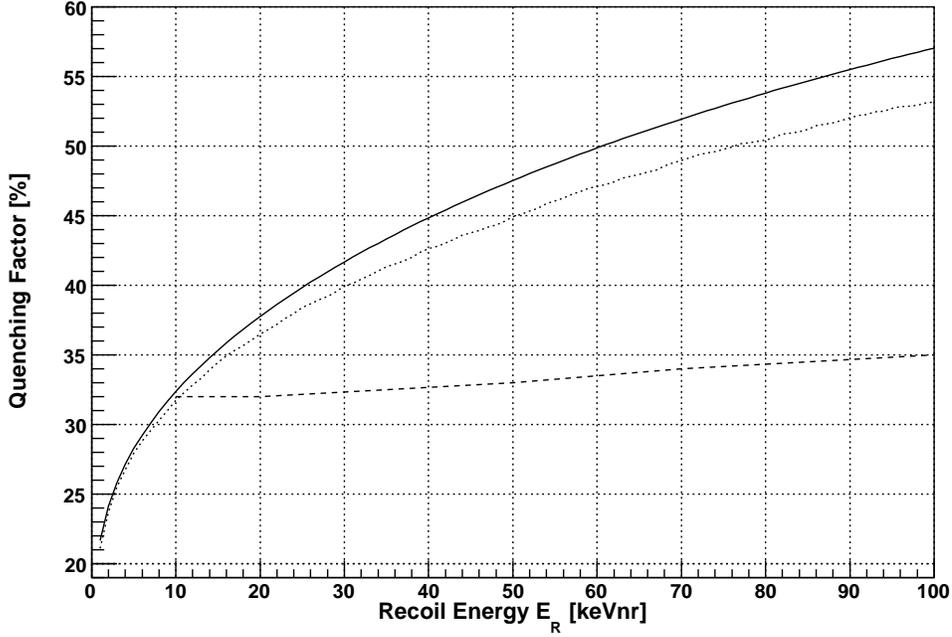}
    \caption{Theoretical curves for the quenching factor of sodium recoils.
      The solid black curve represents the quenching factor of sodium recoils
      in sodium derived from Lindhard
      theory~\cite{Lindhard63_1}~\cite{Lindhard63_2}. The preliminary curve of
      the quenching factor of sodium recoils in NaI(Tl) from~\cite{Hitachi06}
      is illustrated by the dashed line. Finally, the result derived from
      TRIM~\cite{SRIM} for Na recoils in NaI(Tl) is shown by the dotted line.}
    \label{QF_theory}
  \end{center}
\end{figure}

By substituting Eq.~(\ref{epsilon_eq}), Eq.~(\ref{kappa_eq}) and
Eq.~(\ref{ge_eq}) into Eq.~(\ref{QFlind_eq}), the quenching factor can be
expressed as a function of nuclear recoil energy. The theoretical dependence
of the quenching factor for sodium recoils in \isotope[23][11]{Na} is shown in
Figure~\ref{QF_theory}. As \isotope[23][11]{Na} and \isotope[127][53]{I} have
significantly different mass and atomic numbers, the quenching factor of
sodium recoils in sodium, and those in iodine are not similar. In the case of
sodium recoils in iodine, the evaluation of the quenching factor is more
complicated, and Eq.~(\ref{QFlind_eq}) can no longer be used. Lindhard theory
can only approximate the quenching factor in such cases at very low
energies~\cite{Lindhard63_2}.

A key requirement of the Lindhard theory is that electronic and nuclear
collisions can be separated. However, the repulsion between two interacting
nuclei makes part of the parameter range unavailable for transferring energy to
electrons. As a result, the electronic stopping power is suppressed when
$\epsilon\ll 1$, leading to the non-proportionality of
$\left(\frac{d\epsilon}{d\rho}\right)_{\mathrm{e}}$ with $\sqrt{\epsilon}$ in
this energy range~\cite{Tilinin95}. This can be corrected for by replacing
$\xi_{\mathrm{e}}$ in Eq.~(\ref{kappa_eq}) with the function
$\tau\left(\epsilon,\frac{Z_{1}}{Z_{2}}\right)$ given in~\cite{Tilinin95},
where $Z_{1}$ and $Z_{2}$ are the atomic numbers of the penetrating and target
nuclei respectively. The impact of this correction on light nuclei, such as
sodium, is very small, and as such it is not evaluated here.

In semiconductors, the measured quenching factor agrees well with that given by
Eq.~(\ref{QFlind_eq}). For scintillators, however, some degree of quenching
also affects the electronic energy loss by ions (energy loss due to excitation
and ionisation)~\cite{Hitachi05} for high Linear Energy Transfer (LET). The
absolute quenching factor for nuclear recoils $q_{\mathrm{n}}$ given by
Eq.~(\ref{QFlind_eq}) is not the required correction factor to the differential
WIMP-nucleon event rate. The quenching factor of nuclear recoils relative to
those from electron interactions $Q$ can be approximated by~\cite{Hitachi05}:

\begin{equation}\label{QFhitachi_eq}
Q = \frac{q_{\mathrm{n}}q_{\mathrm{e}}}{S_{\gamma}}
\end{equation}
where $q_{\mathrm{e}}$ is the electronic quenching factor (quenching factor for
electronic energy loss of ions) and $S_{\gamma}$ is
the scintillation efficiency for electron recoils. If the quenching factor
for sodium recoils in Na from Eq.~(\ref{QFlind_eq}) is defined as
$q_{\mathrm{n}}$ in Eq.~(\ref{QFhitachi_eq}), then an overestimation for the
quenching factor of sodium recoils in NaI(Tl) will result~\cite{Hitachi06}.

The response of NaI(Tl) to photons is known to be
non-linear~\cite{Aitken67}~\cite{Rooney97} with energy. Therefore, the choice
of gamma source for detector calibration plays some role in the final
quenching factor, as a linear energy distribution is assumed in dark matter
experiments. Using the response curve from~\cite{Aitken67}, $S_{\gamma}$ is
equal to 0.9 for 122~keV gamma-rays. The preliminary theoretical curve of the
quenching factor of Na recoils in NaI(Tl) from~\cite{Hitachi06} is shown in
Figure~\ref{QF_theory}.

The Stopping and Range of Ions in Matter (SRIM) package~\cite{SRIM} simulates
the process of ions impinging onto various target materials. The program
calculates the stopping power and range of ions in matter using a quantum
mechanical treatment of ion-atom collisions. These parameters are used by the
TRansport of Ions in Matter (TRIM) program~\cite{SRIM} to calculate the final
distribution of the ions. All the energy loss mechanisms associated with
ion-atom collisions, such as target damage, sputtering, ionisation and phonon
production, are also evaluated. The quenching factors for various materials
have been simulated with these programs~\cite{Mangiarotti07}, and that for
sodium recoils in NaI(Tl) is determined here using the same technique.

A NaI(Tl) crystal of density 3.67~$\mbox{g/cm}^{3}$ is defined as the solid
target. Sodium ions are given an initial energy (in other words, a recoil
energy) and propagated through the crystal at a normal incidence angle. Recoil
energies are varied between 1~and 100~keV, in 1~keV steps, and 4~000~ions are
simulated at each energy.

The percentage energy loss from the original ions and the resultant recoiling
atoms induced by ion-atom collisions is calculated by TRIM. This is then
subdivided into energy losses from ionisation, vacancies from un-filled holes
left behind after a recoil atom moves from its original site, and phonon
emission. Light emission is a result of ionisation, and hence, the sum of the
percentage energy loss due to ionisation from the original ion and recoiling
atoms is the quenching factor. The mean of these contributions over 4~000
events is evaluated by TRIM, and the results are shown in
Figure~\ref{QF_theory}.

Unlike the prediction of the quenching factor from~\cite{Hitachi06}, the
result from TRIM follows the shape of the Lindhard curve. However, although
they display similar values at low energies, the quenching factor from
Lindhard theory rises faster with increasing energy. At 10~keVnr, all three
results are in good agreement, and it is after this point that they start to
diverge. The most comprehensive treatment to the evaluation of the quenching
factor is that given by~\cite{Hitachi06} (Eq.~(\ref{QFhitachi_eq})).
Therefore, it is reasonable to assume that the measurements of $Q$ will more
closely match the shape of this curve, although their values should lie below
it.

The quenching factor can be measured by inducing nuclear recoils of a known
energy in the target material. In this way, the ratio of measured energy
through electronic energy losses to known recoil energy can be determined. In
the case of NaI(Tl), iodine recoils will also occur. However, it is clear from
Eq.~(\ref{QFlind_eq}) that the degree of quenching for heavy nuclei such as
iodine is significantly greater than that for lighter nuclei such as sodium.
This means that a low energy threshold is required to witness iodine recoils.
As the purpose of this paper is the measurement of Na recoil in NaI(Tl), such
a threshold is not attained, and hence iodine recoils will not be visible.

\section{Experimental apparatus}

Two Sodern GENIE 16 neutron generators are housed within a dedicated neutron
laboratory at the University of Sheffield. The deuterium-deuterium and
deuterium-tritium accelerators produce an isotropic distribution of 2.45~MeV
and 14.0~MeV mono-energetic neutrons respectively. All electronics and data
acquisition equipment are located in the control room, which is isolated from
the experimental hall by 3~ft of concrete shielding. During operation, the beam
is placed into a concrete castle to provide additional shielding.

\begin{figure}
  \begin{center}
    \includegraphics[angle=270,scale=0.7]{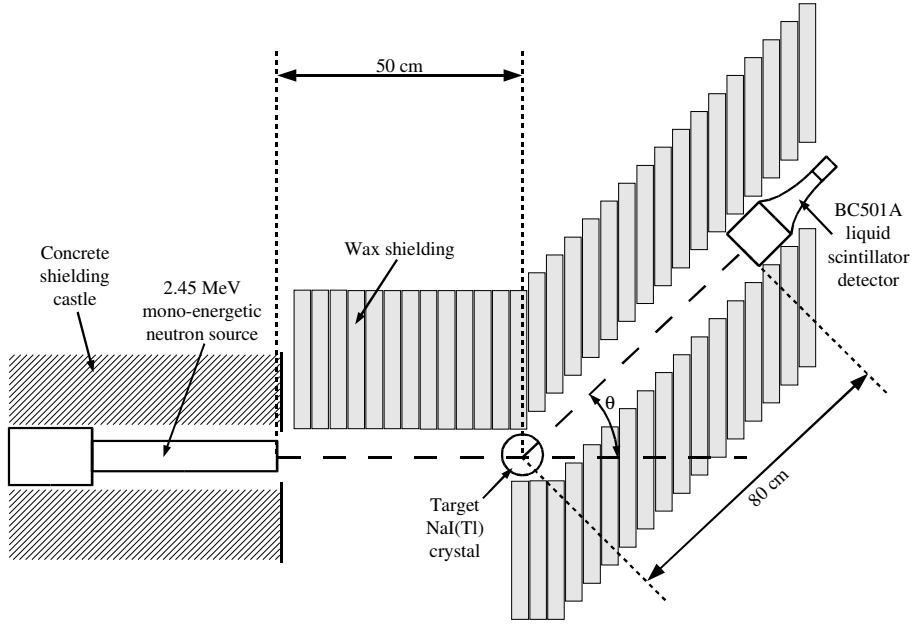}
    \caption{Schematic view of the detector arrangement used to measure
      scintillation from nuclear recoils in NaI(Tl).}
    \label{qf_setup}
  \end{center}
\end{figure}

A schematic view of the detector arrangement is shown in
Figure~\ref{qf_setup}. Only the deuterium-deuterium neutron beam is used for
these measurements. Neutrons of energy 2.45~MeV pass through a hole in the
concrete castle. They travel 50~cm before reaching the centre of the NaI(Tl)
crystal. The energy deposited $E_{R}$ as a function of the scattering angle is
given by:

\begin{equation}\label{nscatter_eq}
E_{R} \approx \frac{2m_{A}E_{\mathrm{n}}m_{\mathrm{n}}}
{(m_{A} + m_{\mathrm{n}})^{2}} \cdot (1 - \cos{\theta})
\end{equation}
where $m_{A}$ is the mass of the target nucleus, $E_{\mathrm{n}}$ is the energy
of incident neutrons, $m_{\mathrm{n}}$ is the mass of the neutron and $\theta$
is the scattering angle. Scattered neutrons are detected by a secondary BC501A
liquid scintillator detector, which is placed at an angle $\theta$ for the
recoil energies of interest $E_{R}$.

NaI(Tl) crystals are hygroscopic and need to be encased within an air-tight
container. The 5~cm diameter, 5.4~cm long, cylindrical NaI(Tl) crystal used
here is encased within a hollow aluminium cylinder of wall thickness 2.5~mm. A
glass window, of thickness 2.5~mm and diameter 5~cm, is optically coupled to
the crystal with silicon oil to improve light collection. The reflection of
light off the inner walls is increased by the wrapping of 1~mm thick PTFE tape
around the crystal. A 3-inch ETL~9265KB photomultiplier tube (PMT)~\cite{ETL}
is optically coupled to the glass window. As the energy of the calibration
source is significantly higher than the nuclear recoil energies in this
experiment, a tapered voltage divider network is chosen and constructed for the
PMT. Such a system reduces space charge effects, which lead to a non-linear
response where high-energy pulses appear smaller than they actually are.

The secondary detector consists of a cylindrical aluminium vessel
of diameter 7.8~cm and height 8.0~cm filled with BC501A liquid scintillator. 
The active volume is viewed by an ETL~9288B
PMT~\cite{ETL} at an operating voltage of -1300~V. 

Only events with a single interaction in the crystal contribute to the recoil
energy $E_{R}$ at a given scattering angle $\theta$ in
Eq.~(\ref{nscatter_eq}). Multiple interactions lead to neutrons depositing a
range of possible energies in the crystal before being detected by the
secondary BC501A detector, thus contributing to the background. Due to the
cylindrical geometry of the crystal, a Monte Carlo simulation is required to
obtain an accurate probability for multiple interactions.

The geometry of the experiment, illustrated in Figure~\ref{qf_setup} is
replicated within the GEANT4 framework~\cite{Agostinelli03}, where 2.45~MeV
neutrons are generated at the face of the neutron source and fired towards the
NaI(Tl) crystal. A total of $10^{8}$~events are generated at scattering angles
associated with 10~and 100~keVnr sodium recoil energies as given by
Eq.~(\ref{nscatter_eq}). Only events that deposit energy in both the crystal
and BC501A detector are recorded. Approximately 0.13\% of 10~keVnr and 0.05\%
of 100~keVnr events satisfy this condition. The reason for this asymmetry is
the non-isotropic cross-section for neutron scattering at higher recoil
energies and for heavier nuclei~\cite{Barschall40}.

\begin{figure}
  \begin{center}
    $\begin{array}{c@{\hspace{5mm}}c}
      \includegraphics[scale=0.35]{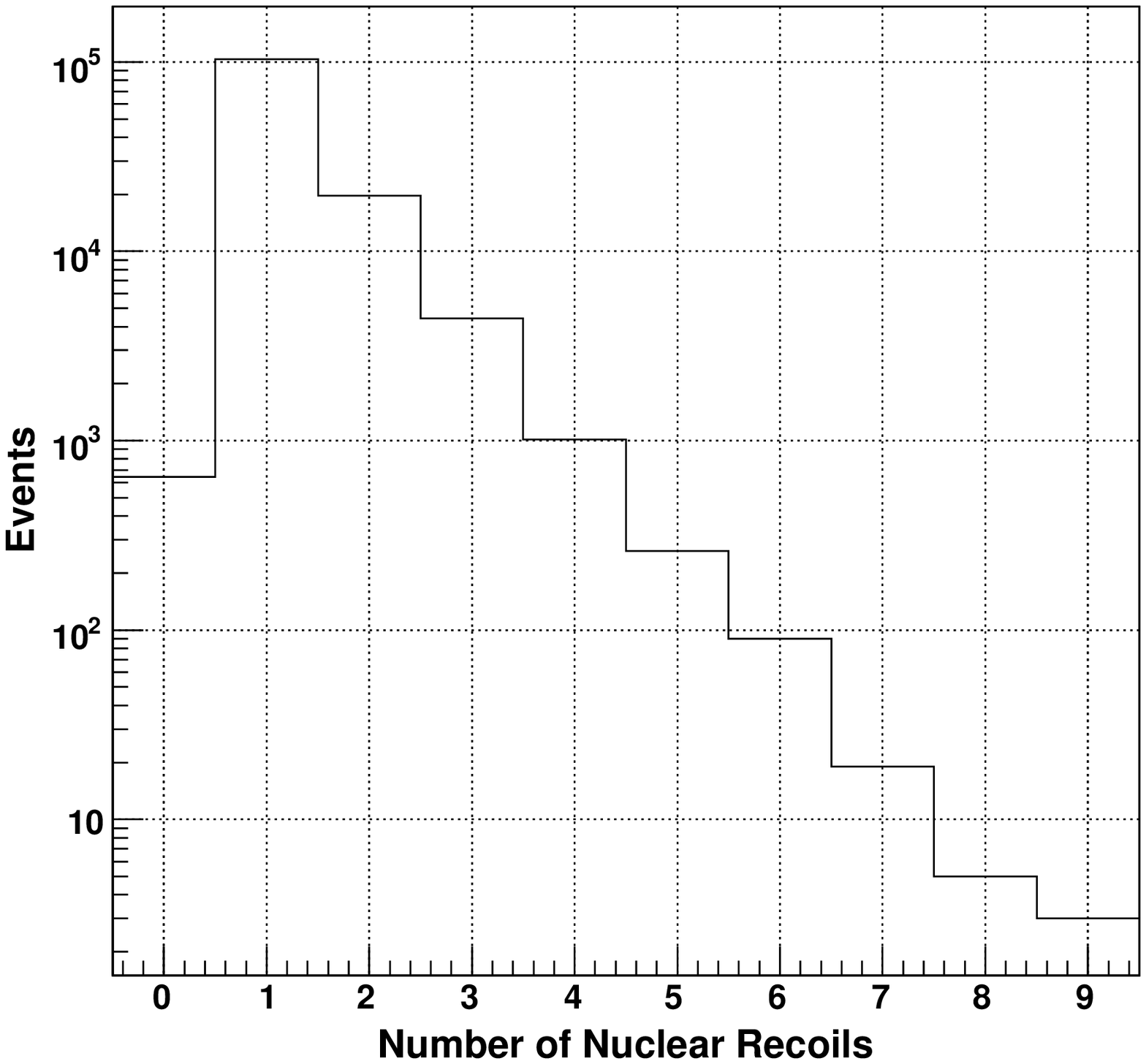} &
      \includegraphics[scale=0.35]{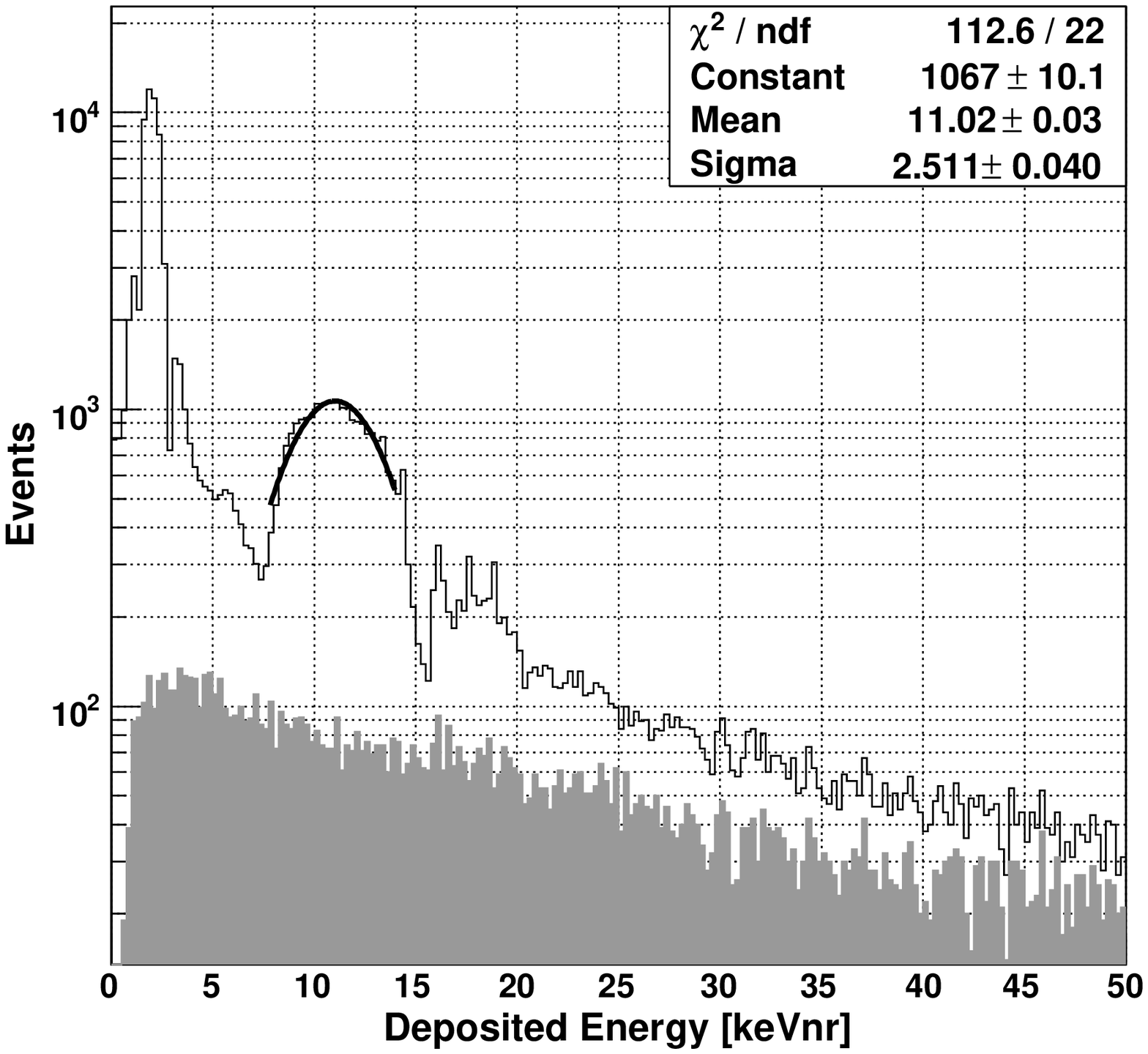} \\
      \mbox{(a)} & \mbox{(b)}
    \end{array}$
    \caption{Simulated distributions of: (a)~number of nuclear recoils in NaI
      per neutron; and (b)~total energy deposited in NaI(Tl) crystal. A
      significant proportion of events do result in multiple scattering, as
      shown in~(a). However, the deposited energy spectrum for events as a
      result of two or more nuclear recoils, represented by the shaded area
      in~(b), is featureless. This implies that their contribution to the
      background should not interfere with the signal peak position at
      approximately 10~keVnr. The peak at approximately 2~keVnr in~(b) is from
      iodine recoils, which is not visible in real data at this scattering
      angle due to the higher energy threshold. Features either side of the
      recoil peaks in~(b) are due to neutrons scattering off nuclei within the
      wax shielding before entering the secondary detector, and those that
      escape through gaps between the BC501A cylinder and wax walls. The
      contribution to background from these interactions at other deposited
      energies is featureless.}
    \label{nrg_sim}
  \end{center}
\end{figure}

Simulation results at 10~keVnr nuclear recoil energy are shown in
Figure~\ref{nrg_sim}. Although a significant proportion of events undergo
multiple scattering in the crystal, the deposited energy from these
interactions, represented by the shaded histogram, is featureless in
comparison with the total recoil energy spectrum. Therefore, there is no
preferential energy deposition, and the number of multiple interactions should
make no difference to the final result.

The inclusion of nuclear recoils off iodine nuclei also results in a low energy
peak at approximately 2~keVnr from single scattered neutrons as shown in
Figure~\ref{nrg_sim}(b). From Eq.~(\ref{nscatter_eq}), the change in energy
with scattering angle is far more pronounced for lighter nuclei, and as iodine
has a significantly higher mass number than sodium, such a result is expected.
For the reasons outlined in Section~\ref{theory_sec}, such a peak would not be
visible in the measured data, and will not interfere with the sodium peak.

\begin{figure}
  \begin{center}
    \includegraphics[scale=0.6]{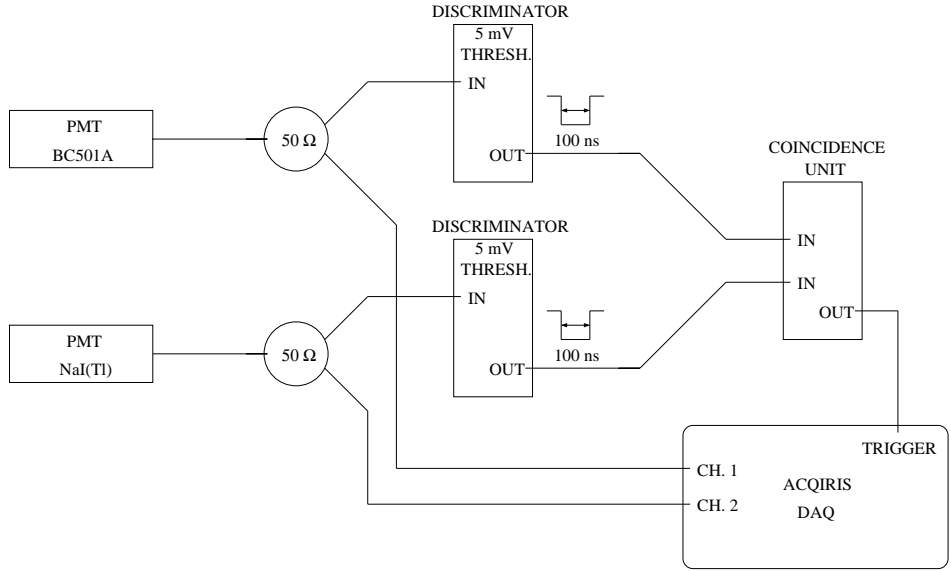}
    \caption{Hardware trigger electronics for the quenching factor experiment.
      Analogue photomultiplier signals from the BC501A detector and NaI(Tl)
      crystal are split with a 50~$\Omega$ power divider, and sent to a
      discriminator and an input channel on the DAQ.
      The discriminator is set at a threshold of 5~mV, and a 100~ns wide NIM
      pulse is sent to a 2-fold coincidence unit. If the signals are
      coincident, a NIM pulse provides the external trigger to the DAQ.}
    \label{nbeam_elec}
  \end{center}
\end{figure}

The configuration of electronics for the hardware trigger is shown in
Figure~\ref{nbeam_elec}. Analogue photomultiplier signals from the NaI(Tl)
crystal and BC501A detector are split with a 50~$\Omega$ power divider. The
signal from each PMT is then sent simultaneously to a discriminator set at a
threshold of 5~mV, and a channel of the data acquisition system (DAQ). The
hardware trigger is two signals coincident within a 100~ns time window. The
analogue pulses are converted to digitised waveforms by an 8-bit, 2-channel
Acqiris digitiser~\cite{Acqiris} with a 500~MHz sampling rate. Data
acquisition software running on a linux computer, similar to that used by the
ZEPLIN-II experiment~\cite{Alner08}, reads out the digitised waveforms and
writes them to disk.

An analysis program reads the binary data output of the digitiser. The program
goes through each event, extracting the amplitude at each 2~ns sampling point
and placing the values into an array. To assign a baseline, the mean and
standard deviation of the first 200~ns of a waveform are determined. This
process is then repeated over the same time window, excluding bins of
amplitude greater than three standard deviations from the mean. The baseline
is calculated in this manner on an event-by-event basis, and this procedure
results in an improvement to its estimation.

An event viewer is implemented within the ROOT framework~\cite{Brun97}.
Waveform parameters are extracted and stored in a ROOT tree for later analysis.
The total pulse area, which is proportional to the deposited energy, is the sum
of digitised bin contents within a range:

\begin{equation}\label{pulsearea_eq}
\sum^{s_{2}}_{i = s_{1}}{V_{i}(t)\Delta t} = \int^{t_{2}}_{t_{1}}{V(t)dt}
\end{equation}
where $s_{1}$ is the first and $s_{2}$ the second sampling point over which
the summation is performed. The value of $s_{1}$ is the first point at which
the pulse reaches 10\% of its maximum amplitude. The amplitude of each bin $i$
is denoted by $V_{i}(t)$, and with a 500~MHz sampling rate, $\Delta t = 2$~ns.
The start $t_{1}$ and stop $t_{2}$ times are defined as
$t_{1,2} = s_{1,2}\Delta t$.

\section{Measurements}

\subsection{Gamma-ray calibration}

\begin{figure}
  \begin{center}
    $\begin{array}{c@{\hspace{5mm}}c}
      \includegraphics[scale=0.35]{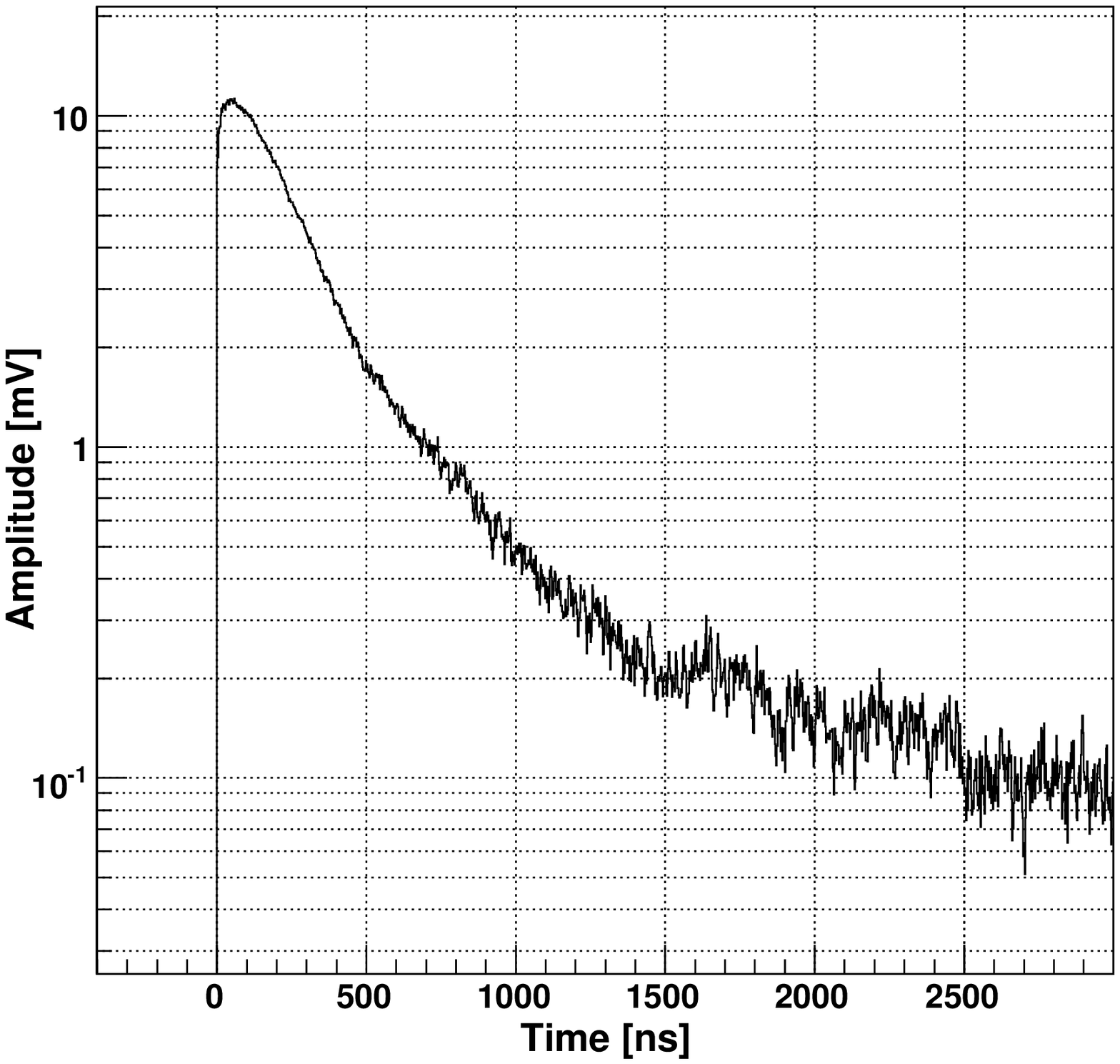} &
      \includegraphics[scale=0.35]{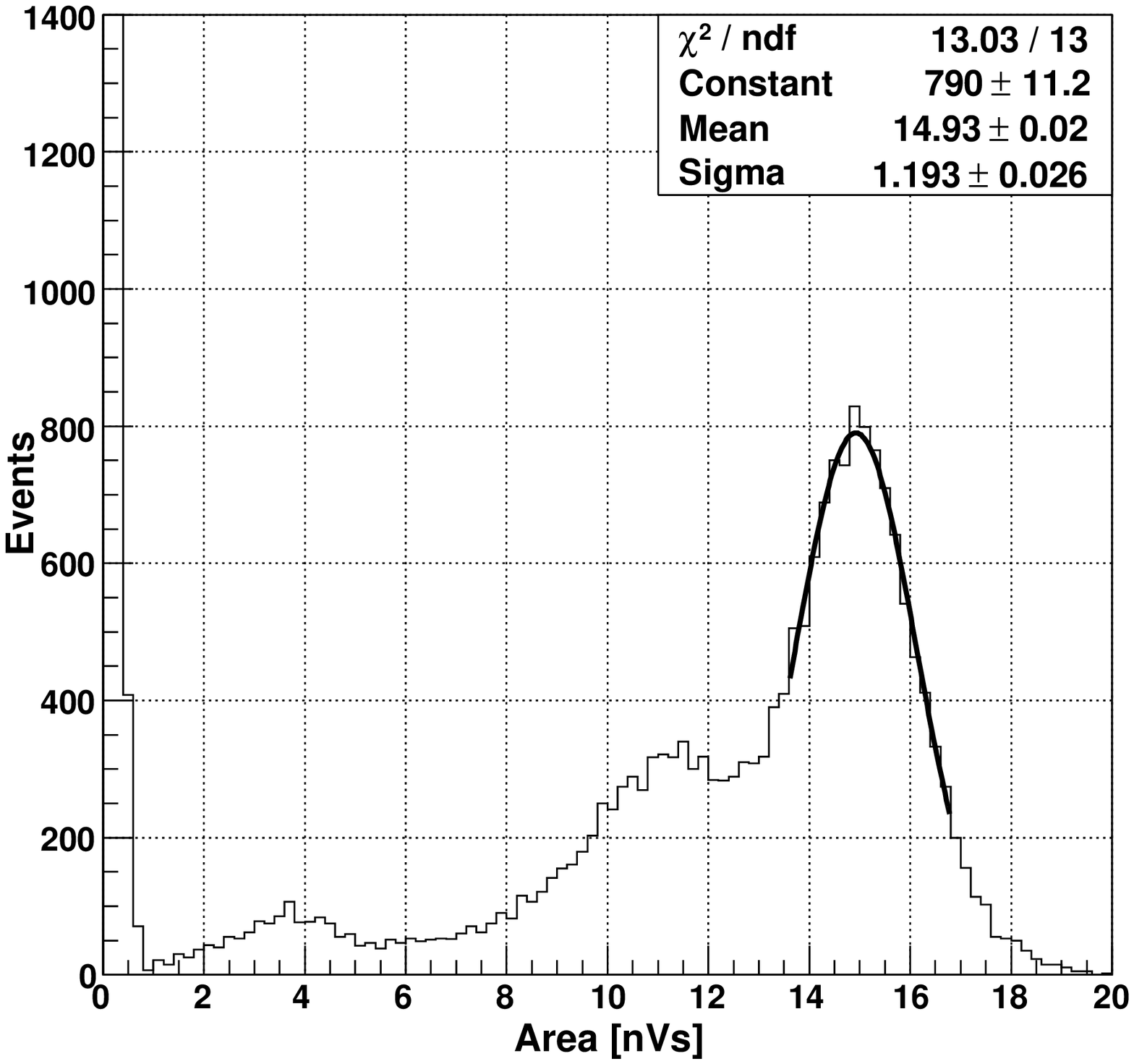} \\
      \mbox{(a)} & \mbox{(b)}
    \end{array}$
    \caption{(a)~Pulse shape for an event induced by 30~keV X-rays 
    from \isotope[129]{I} source. Pulses that contribute
      to the 30~keV peak are extracted and their amplitudes are summed. The
      pulse is normalised to 1~event. (b)~Calibration of NaI(Tl) crystal
      detector with 122~keV gamma line from \isotope[57]{Co}. Three clear
      peaks are visible corresponding to X-rays, X-ray escape peak and total
      absorption at approximately 3.5, 11.5~and 15~nVs respectively.}
    \label{y_nai}
  \end{center}
\end{figure}

A constant value for $t_{2}$ in Eq.~(\ref{pulsearea_eq}) needs to be defined
to obtain the area under NaI(Tl) pulses. It is difficult to determine $t_{2}$
from a single pulse, due to the sparse distribution of photons in the tail
region. Instead, a scintillation pulse is built from the sum of 10~000
pulses detected when the crystal is irradiated by 30 keV X-rays from
\isotope[129]{I} source, as shown in Figure~\ref{y_nai}. The amplitudes are
normalised to reproduce the mean shape of one pulse.

Photons from scintillation light continue well beyond 3~$\mu$s. However,
electronic noise in the tail region makes it difficult to choose a relatively
large value for $t_{2}$. A compromise of 2~$\mu$s after the start of the pulse
is used, which is equivalent to about 92\% of the total waveform.

The crystal is exposed to gamma-rays from a variety of radioactive sources,
between 30~keV (X-rays from \isotope[129]{I}) and 662~keV
(\isotope[137]{Cs} $\gamma$-rays). A decrease in photon response is observed at
the iodine K-shell absorption edge at 33.2~keV, consistent with other
studies~\cite{Aitken67}~\cite{Rooney97}. Therefore, determination of the
energy scale must be performed in a region where a linear response is
observed. Calibration is performed with the 122~keV gamma line from a
\isotope[57]{Co} source to establish an electron equivalent energy scale
(labelled keVee as opposed to keVnr for nuclear recoil energies), as shown in
Figure~\ref{y_nai}. A light yield of 5.1~photoelectrons/keV is found. The
procedure is repeated approximately every 3~hours to analyse any drift in the
light yield, and if significant degradation is witnessed the crystal is
recoupled to the PMT.

An attenuation coefficient of around 1.01~$\mbox{cm}^{2}$/g for 122~keV
\isotope[57]{Co} gamma-rays traversing NaI(Tl) translates to a mean free path
length of 2.7~mm. Therefore, most interactions will occur near the surface of
the crystal, and hence may be affected by defects. To check for deformities,
the light yield at $30^{\circ}$ angles around the crystal is checked. All but
one point lie within one standard deviation of the mean light yield.
Therefore, the use of \isotope[57]{Co} for calibration is acceptable, as no
major surface defects are present.

A suitable full scale (range) over which to digitise the signal needs to be
chosen. Using the Lindhard curve for sodium recoils in Na from
Figure~\ref{QF_theory}, a 50~keV nuclear recoil will be quenched by 48\%,
resulting in a 25~keV electron equivalent pulse. This is equivalent to a pulse
close to that from a 30~keV X-rays from \isotope[129]{I}. 
With an amplitude of just over 11~mV, a range of
50~mV is adequate for this experiment.

Low level cuts include removing pulses that saturate the digitiser and are not
in coincidence. The latter are caused when the DAQ triggers on the end of an
event.

\subsection{Event selection by pulse shape discrimination in BC501A}

A secondary detector is required to identify neutrons that scatter off the
target nuclei at the nuclear recoil energies given by Eq.~(\ref{nscatter_eq}).
As the main background is from gamma-rays, a detector material with a high
discrimination power is a major requirement. Enhanced emission of the slow
component and a high hydrogen-to-carbon ratio make the BICRON Corporation
BC501A organic liquid scintillator
($\mbox{C}_{6}\mbox{H}_{4}(\mbox{CH}_{4})_{2}$), equivalent to Nuclear
Enterprises NE213, well-suited for this purpose.

\begin{figure}
  \begin{center}
    \includegraphics[scale=0.7]{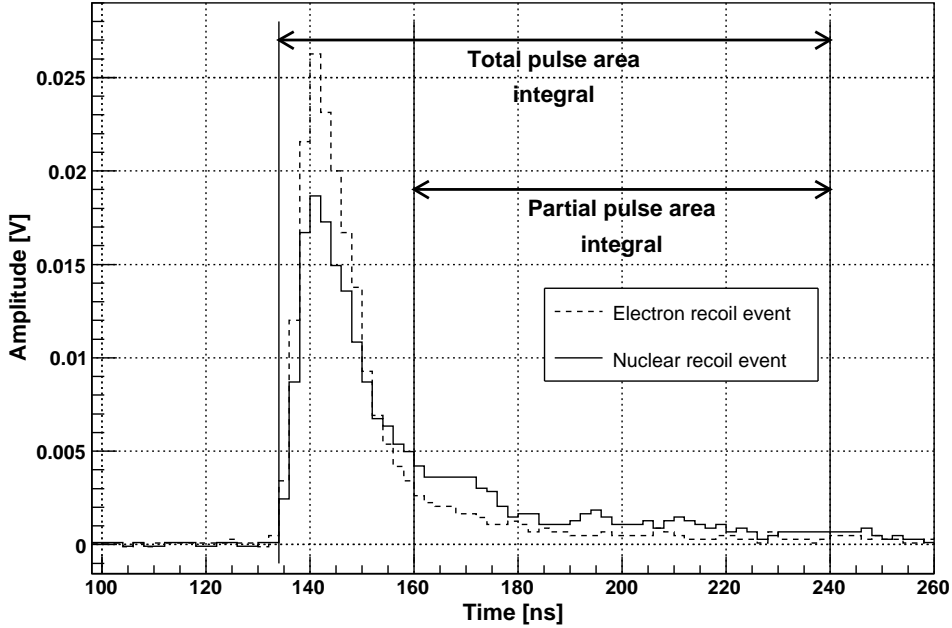}
    \caption{Typical pulses from 600~keVee nuclear and electron recoils in the
      BC501A detector. Energy scale calibration is performed with the 662~keV
      gamma line from a \isotope[137]{Cs} source. The total pulse area $A$ is
      the sum of the bins that lie between the start time and 100~ns after the
      maximum peak position. The partial pulse area $P$ is the sum of those
      that lie between 20~ns after the peak position and the end of the
      waveform.}
    \label{bc501a_waveforms}
  \end{center}
\end{figure}

As with the NaI(Tl) pulses described previously, a suitable value for $t_{2}$
in Eq.~(\ref{pulsearea_eq}) needs to determined. From the typical 600~keVee
pulses in Figure~\ref{bc501a_waveforms}, a value of 100~ns after the position
of the maximum bin is defined as the end of the waveform.

The intensity of these pulses $I(t)$ can be written to good approximation as a
function of four exponentials~\cite{Marrone02}:

\begin{equation}\label{bcpulsefit_eq}
I(t) \approx A \left(e^{-\frac{(t - t_{0})}{\tau_{e}}} -
e^{-\frac{(t - t_{0})}{\tau_{1}}} +
\frac{B}{A} e^{-\frac{(t - t_{0})}{\tau_{e}}} -
\frac{B}{A} e^{-\frac{(t - t_{0})}{\tau_{2}}}\right)
\end{equation}
where $\tau_{e}$ is the RC time constant of the data acquisition electronics,
$\tau_{1}$ and $\tau_{2}$ are the decay time constants of the fast and slow
components respectively, and $A$ and $B$ are their respective intensities. Due
to the loose definition of the pulse start time, an additional parameter for
time reference $t_{0}$ is defined.

The ratio $\left\vert\frac{B}{A}\right\vert$ in Eq.~(\ref{bcpulsefit_eq})
provides a measure of the discrimination power, making use of the
characteristic enhanced emission of the slow component in
BC501A~\cite{Marrone02}. The fitting of each pulse is a time consuming
procedure due to the six free parameters in Eq.~(\ref{bcpulsefit_eq}). The time
taken for a fit to successfully converge can be improved by restricting
parameters, or deriving average values for some and fixing
them~\cite{Marrone02}. However, if the discriminating factor is the ratio of
the intensities of the slow to fast components, the same should hold true for
the ratio of the slow component to the total intensity of the pulse as
$B\ll A$.

Therefore, by integrating over the tail of a pulse and dividing by the total
area, neutron and gamma events are separated. The ratio of partial pulse area
in the tail $P_{\mathrm{nr}}$ to total pulse area $A_{\mathrm{nr}}$ will be
closer to unity when compared with electron recoils of the same energy
$\frac{P_{\mathrm{er}}}{A_{\mathrm{er}}}$. This can be inferred
from~\cite{Marrone02}, where the $\left\vert\frac{B}{A}\right\vert$ ratio is
smaller for electron recoils. In other words:

\begin{equation}\label{bcpsd_eq}
\frac{P_{\mathrm{nr}}}{A_{\mathrm{nr}}} >
\frac{P_{\mathrm{er}}}{A_{\mathrm{er}}}
\end{equation}

The discrimination technique defined in Eq.~(\ref{bcpsd_eq}) is tested by
exposing the BC501A detector to 662~keV gamma-rays from a \isotope[137]{Cs}
source and 2.45~MeV neutrons from the deuterium-deuterium beam. The hardware
trigger is identical to that shown in Figure~\ref{nbeam_elec}, with the
exception of the NIM pulse from the discriminator acting as the external
trigger to the digitiser. Offline saturation cuts are performed on the recorded
data.

\begin{figure}
  \begin{center}
    $\begin{array}{c@{\hspace{5mm}}c}
      \includegraphics[scale=0.35]{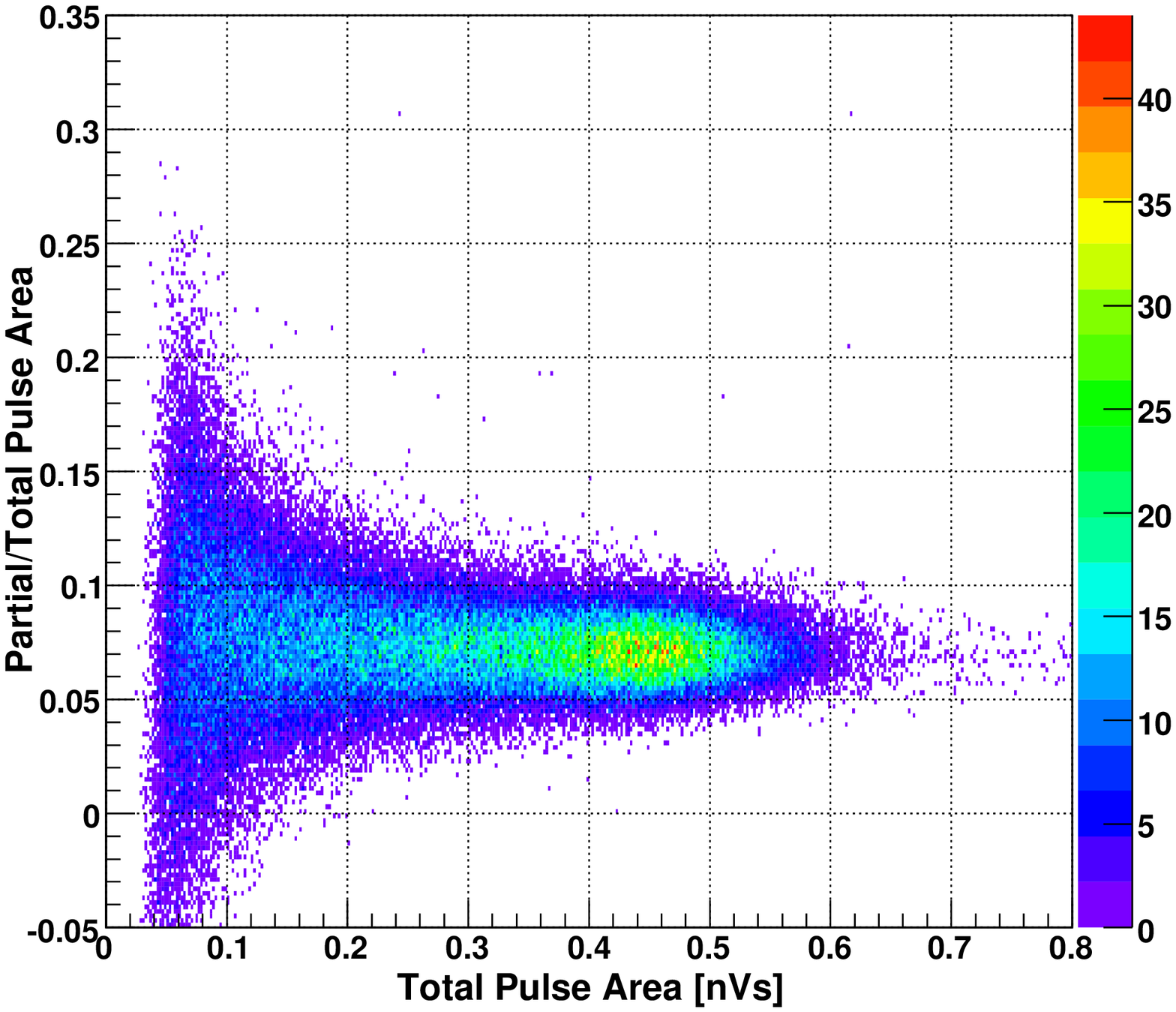} &
      \includegraphics[scale=0.35]{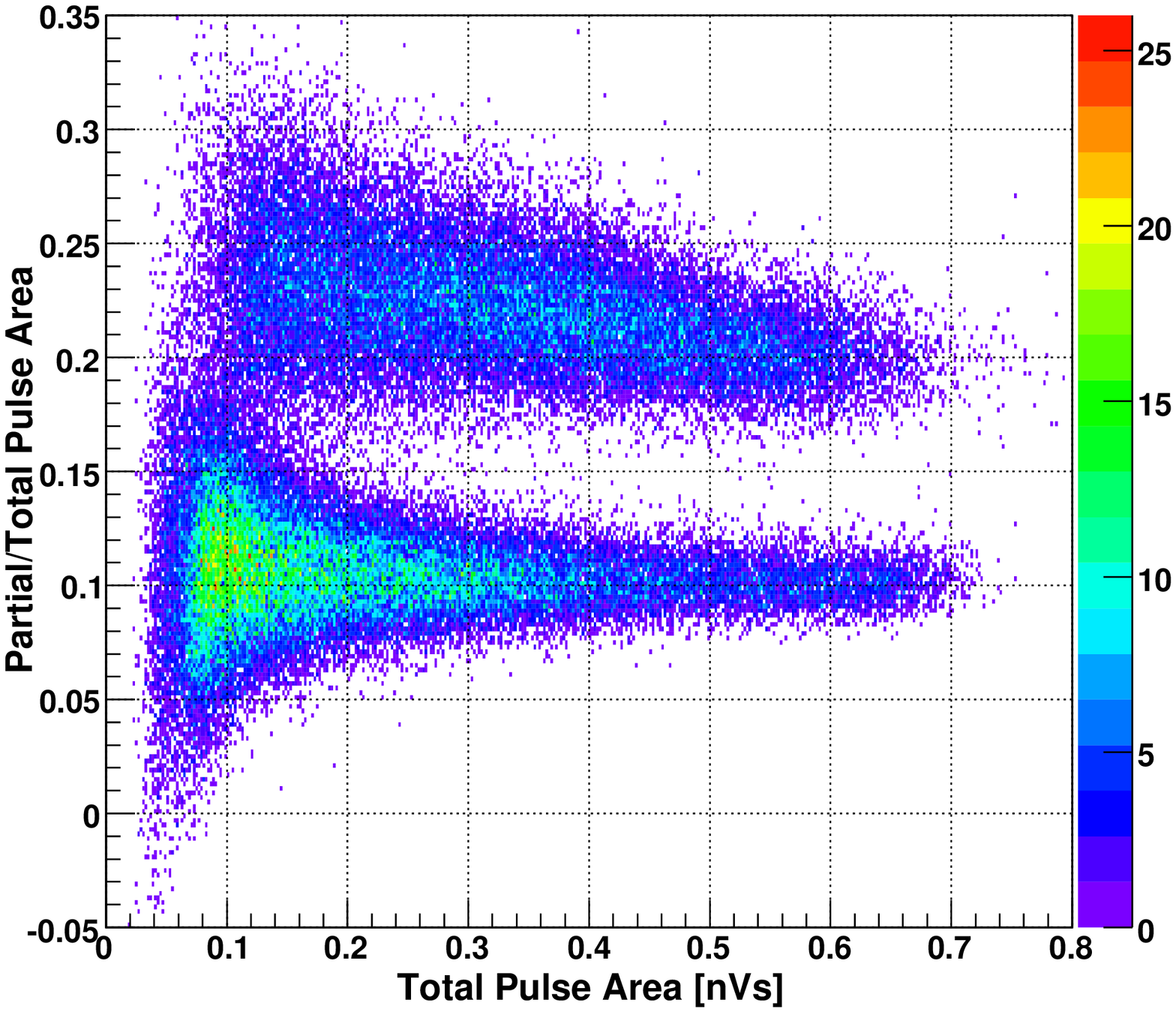} \\
      \mbox{(a)} & \mbox{(b)}
    \end{array}$
    \caption{Scatter plots of partial-to-total pulse area ratio against total
      pulse area in the BC501A detector for: (a)~gamma-rays from a
      \isotope[137]{Cs} source; and (b)~neutrons from the 2.45~MeV
      deuterium-deuterium neutron beam and background gamma-rays. The 662~keV
      gamma line is visible in (a) at approximately 0.45~nVs total pulse area.
      An upper band of neutron energies is evident in (b). Lower limit of
      30~ns after the maximum peak position is implemented for the partial
      pulse area integrals.}
    \label{bc501a_psd}
  \end{center}
\end{figure}

The results are shown in Figure~\ref{bc501a_psd}. A lower limit for the
partial pulse area integral of 30~ns after the maximum peak position is
chosen. The presence of an upper horizontal band of neutrons is seen in the
data from the neutron run. This band is absent from the gamma-ray data. A large
gamma background is apparent in data from the 2.45~MeV deuterium-deuterium
beam, emphasising the need for good neutron-gamma discrimination.

By changing the pulse partial area integration boundaries, it may be possible
to increase the resolution between neutron and gamma events, and hence decrease
the energy threshold for discrimination. The start of the tail is varied
between 10~and 50~ns after the maximum peak position, in stages of 10~ns. One
dimensional histograms of the partial-to-total pulse area ratio result in two
peaks. By fitting Gaussian functions to these peaks, a figure of merit $M$ is
used to quantify the neutron-gamma discrimination power:

\begin{equation}\label{mfact_eq}
M = \frac{(\bar{x}_{\mathrm{n}} - \bar{x}_{\gamma})}
{(\sigma_{\mathrm{n}} + \sigma_{\gamma})}
\end{equation}
where $\bar{x}_{\mathrm{n}}$ and $\bar{x}_{\gamma}$ are the mean positions of
the neutron and gamma peaks respectively. The full width half maximum of the
neutron and gamma peaks are given by $\sigma_{\mathrm{n}}$ and
$\sigma_{\gamma}$ respectively. Applying Eq.~(\ref{mfact_eq}), $M$-factors are
calculated as shown in Table~\ref{mfact_tab}. A lower limit on partial pulse
area integral of 20~ns after the position of the peak provides the best
resolution power.

\begin{table}
\begin{center}
\begin{tabular}{ccccc}
\hline
Lower limit of & & Electron recoil & Nuclear recoil &
\multirow{2}{*}{$M$} \\
integration [ns] & & peak & peak & \\
\hline\hline
\multirow{2}{*}{10} & $\bar{x}$ & $0.2990 \pm 0.0002$ & $0.4518 \pm 0.0003$ &
\multirow{2}{*}{$2.15 \pm 0.02$} \\
& $\sigma$ & $0.0324 \pm 0.0003$ & $0.0386 \pm 0.0005$ & \\
\hline
\multirow{2}{*}{20} & $\bar{x}$ & $0.1555 \pm 0.0001$ & $0.2961 \pm 0.0003$ &
\multirow{2}{*}{$2.97 \pm 0.03$} \\
& $\sigma$ & $0.0189 \pm 0.0002$ & $0.0284 \pm 0.0004$ & \\
\hline
\multirow{2}{*}{30} & $\bar{x}$ & $0.1030 \pm 0.0001$ & $0.2155 \pm 0.0003$ &
\multirow{2}{*}{$2.90 \pm 0.03$} \\
& $\sigma$ & $0.0149 \pm 0.0002$ & $0.0239 \pm 0.0003$ & \\
\hline
\multirow{2}{*}{40} & $\bar{x}$ & $0.0715 \pm 0.0001$ & $0.1596 \pm 0.0002$ &
\multirow{2}{*}{$2.70 \pm 0.03$} \\
& $\sigma$ & $0.0125 \pm 0.0002$ & $0.0201 \pm 0.0003$ & \\
\hline
\multirow{2}{*}{50} & $\bar{x}$ & $0.0482 \pm 0.0001$ & $0.1161 \pm 0.0002$ &
\multirow{2}{*}{$2.49 \pm 0.04$} \\
& $\sigma$ & $0.0108 \pm 0.0001$ & $0.0165 \pm 0.0004$ & \\
\hline
\end{tabular}
\caption{Effects of different lower limits for partial pulse integration
(column 1) on the peak positions for BC501A pulses (columns 2 and 3) and
figure of merit as defined by Eq.~4.3 (column 4).}
\label{mfact_tab}
\end{center}
\end{table}

\begin{figure}
  \begin{center}
    $\begin{array}{c@{\hspace{5mm}}c}
      \includegraphics[scale=0.35]{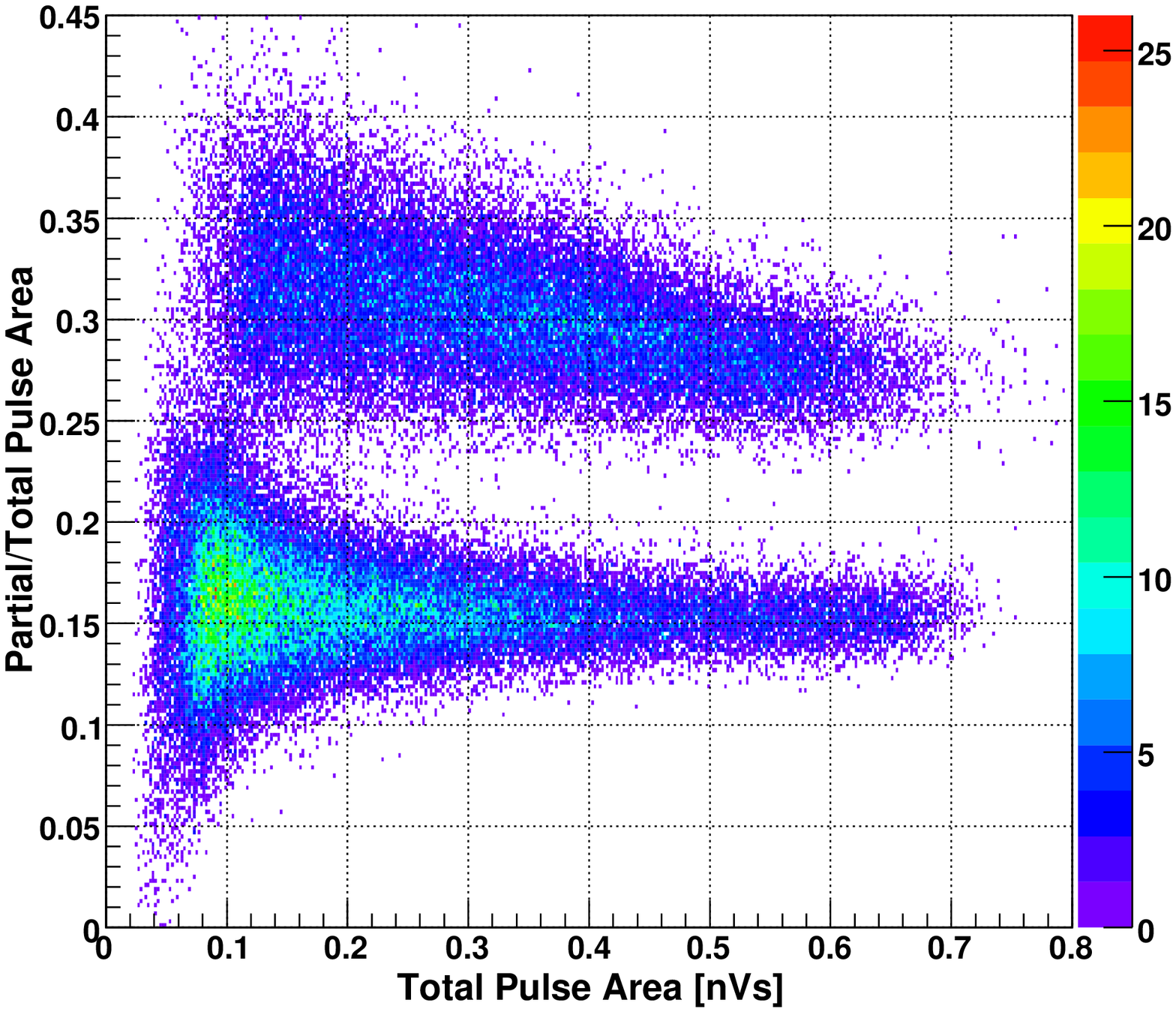} &
      \includegraphics[scale=0.35]{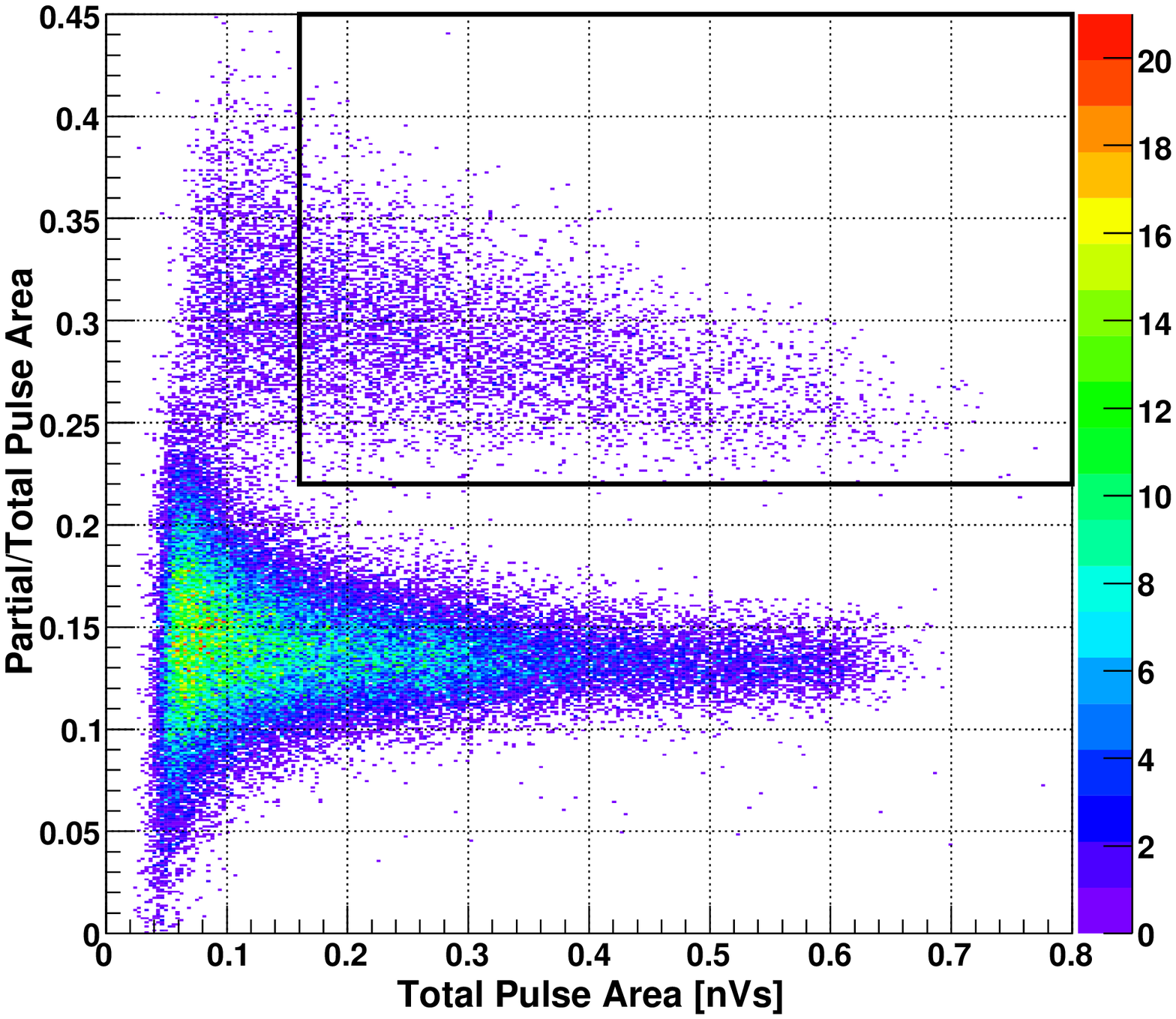} \\
      \mbox{(a)} & \mbox{(b)}
    \end{array}$
    \caption{Implementation of pulse shape discrimination cut in the BC501A
      detector for neutrons from the 2.45~MeV deuterium-deuterium neutron beam.
      The best separation between gammas and neutrons occurs when the lower
      limit for partial pulse integral is set to 20~ns after the maximum peak
      position, as shown in (a). The implementation of this cut is shown for
      the data taken at 10~keVnr in (b), where events that lie within the black
      box are accepted. A minimum energy threshold of 280~keVee is attained.}
    \label{bc501a_cuts}
  \end{center}
\end{figure}

Scintillation light from nuclear recoils is quenched in all materials. BC501A
is not an exception, and the quenching factor has been found to be
non-linear~\cite{Lee98}. An electron equivalent scale is established by
calibrating the BC501A detector with the 662~keV gamma line from
\isotope[137]{Cs}. Cutting at total pulse areas greater than 16~nVs, as shown
in Figure~\ref{bc501a_cuts}, yields a minimum energy threshold of 280~keVee.
Using the non-linear function for the proton quenching factor given
by~\cite{Lee98}, this corresponds to a proton recoil energy of 910~keVnr.
Reducing the threshold further does not affect the resulting quenching
factor.

\subsection{Event selection by time of flight}

As the rest mass energy of incident neutrons is far greater than their kinetic
energy of 2.45~MeV, they are non-relativistic. With reference to
Figure~\ref{qf_setup}, after interacting with the crystal, the non-relativistic
neutrons will take longer to reach the secondary BC501A detector than
gamma-rays. This time of flight $t$ can be quantified with:

\begin{equation}\label{tof_eq}
t\,\mbox{[ns]} \approx 3.336 \cdot s\,\mbox{[m]}
\sqrt{\frac{m_\mathrm{n}\,\mbox{[MeV]}}{2E_{\mathrm{n}}\,\mbox{[MeV]}}}
\end{equation}
where $s$ is the distance travelled by the neutron. As
$E_{R}\ll E_{\mathrm{n}}$, the recoil energy is ignored to good approximation
in Eq.~(\ref{tof_eq}). For $E_{\mathrm{n}} = 2.45$~MeV, Eq.~(\ref{tof_eq})
yields a value of 38~ns for the time of flight.

\begin{figure}
  \begin{center}
    $\begin{array}{c@{\hspace{5mm}}c}
      \includegraphics[scale=0.35]{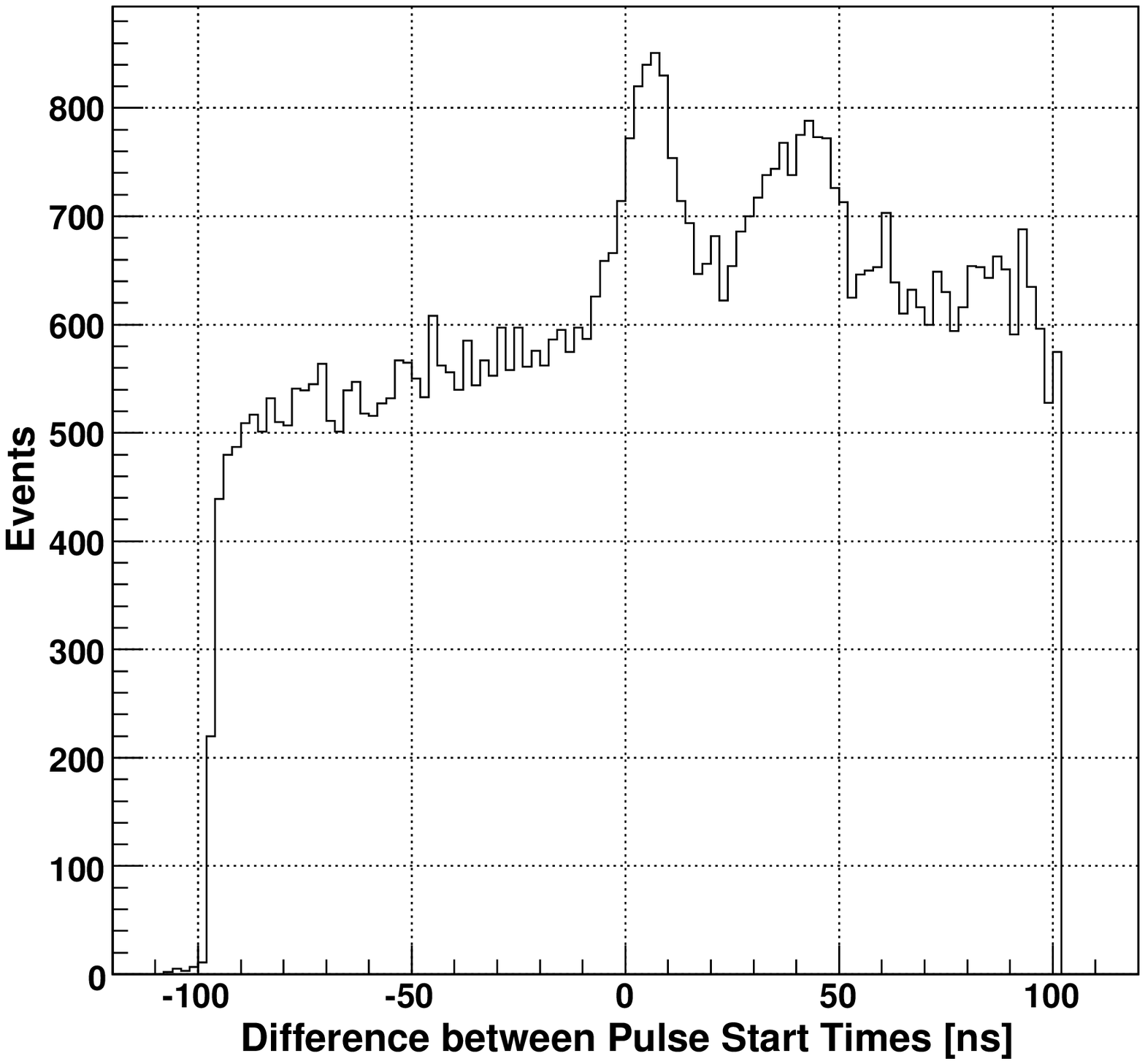} &
      \includegraphics[scale=0.35]{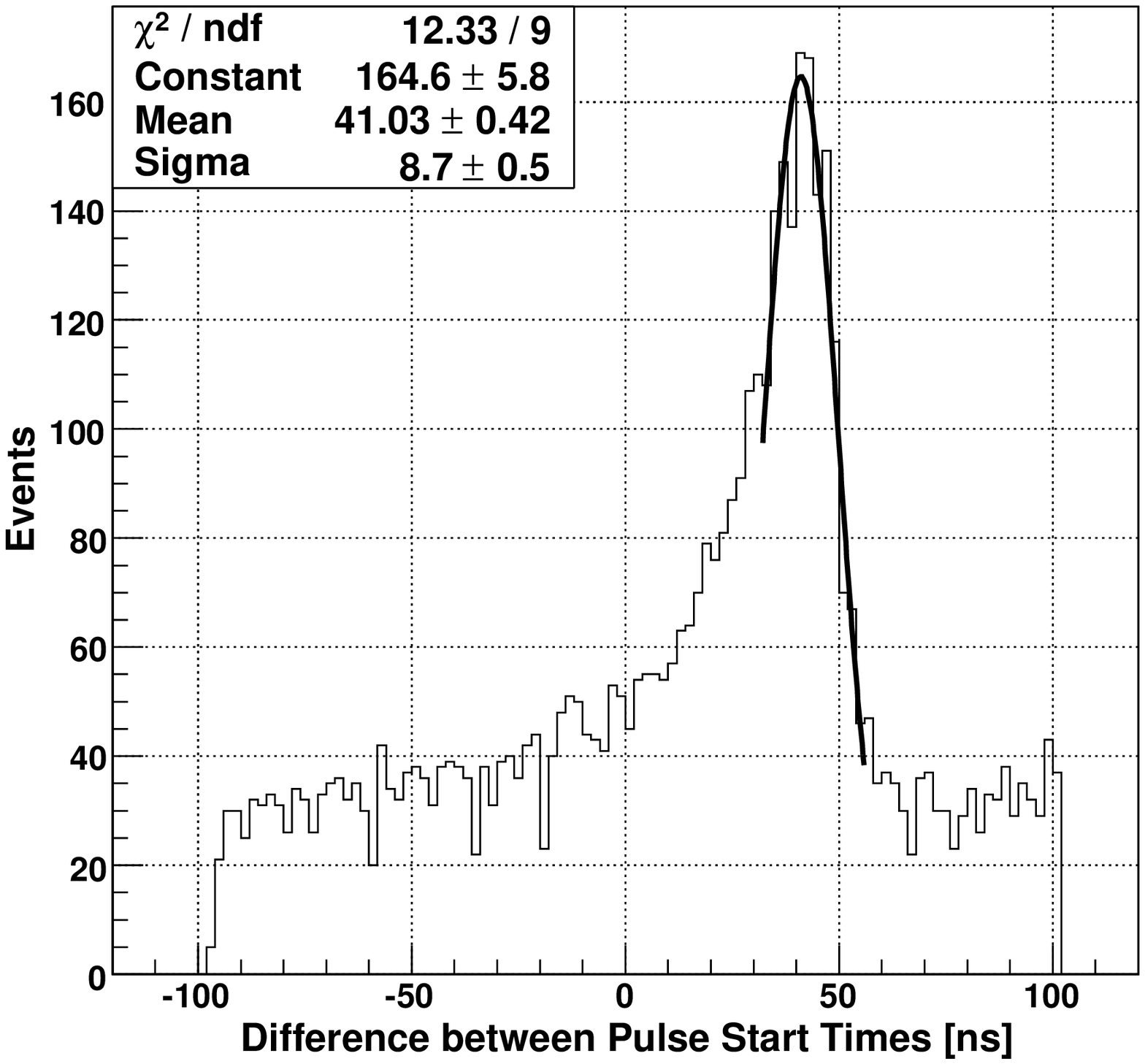} \\
      \mbox{(a)} & \mbox{(b)}
    \end{array}$
    \caption{Time of flight distributions between NaI(Tl) crystal and BC501A
      detector for data taken at 10~keVnr: (a)~before; and (b)~after performing
      the cut on pulse shape discrimination in BC501A. Gamma and neutron peaks
      at approximately 0~and 40~ns are visible in (a). After the cut on pulse
      shape is implemented, it becomes possible to fit a Gaussian function to
      the neutron peak in (b).}
    \label{tof_plots}
  \end{center}
\end{figure}

\begin{figure}
  \begin{center}
    $\begin{array}{c@{\hspace{5mm}}c}
      \includegraphics[scale=0.35]{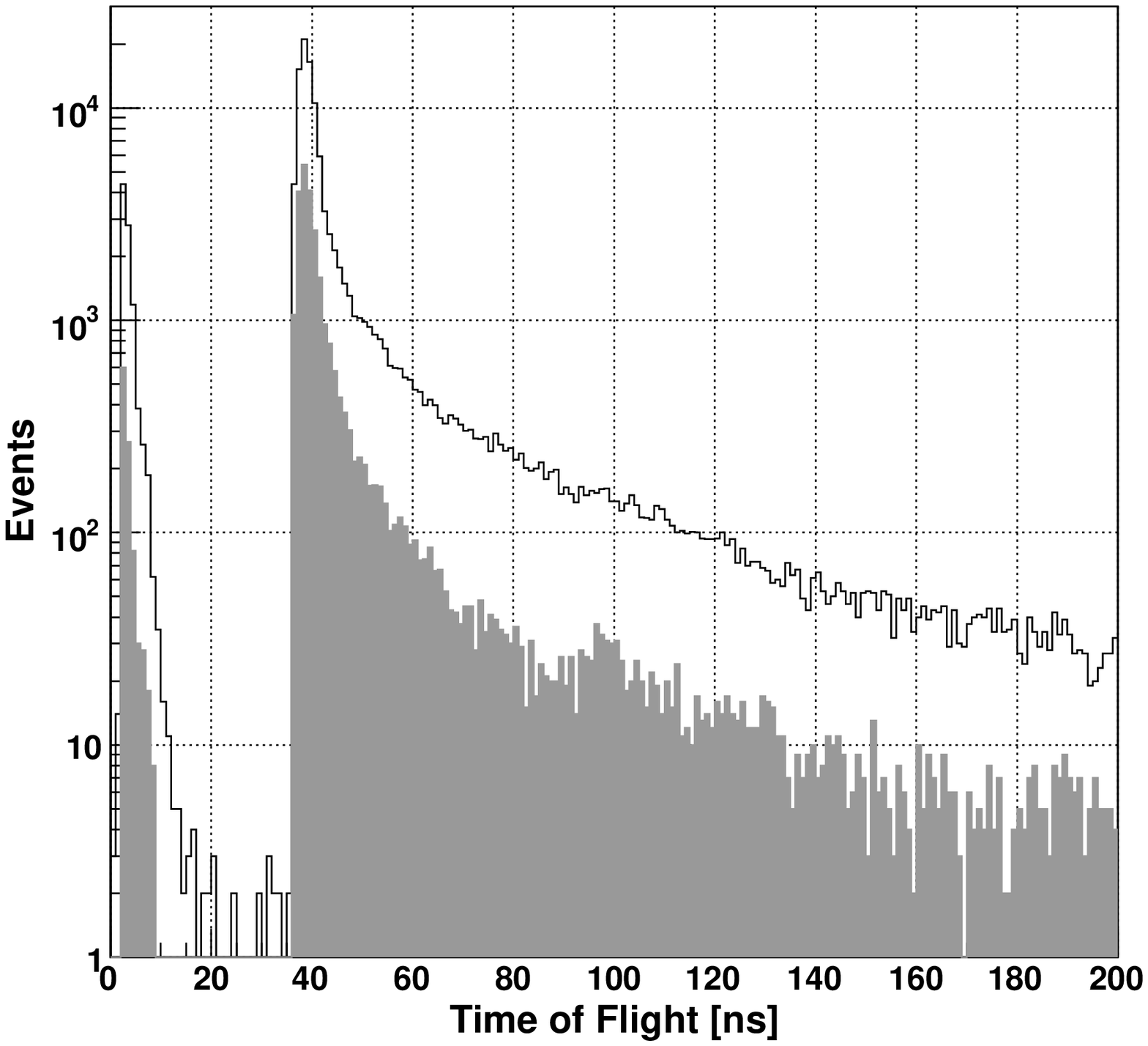} &
      \includegraphics[scale=0.35]{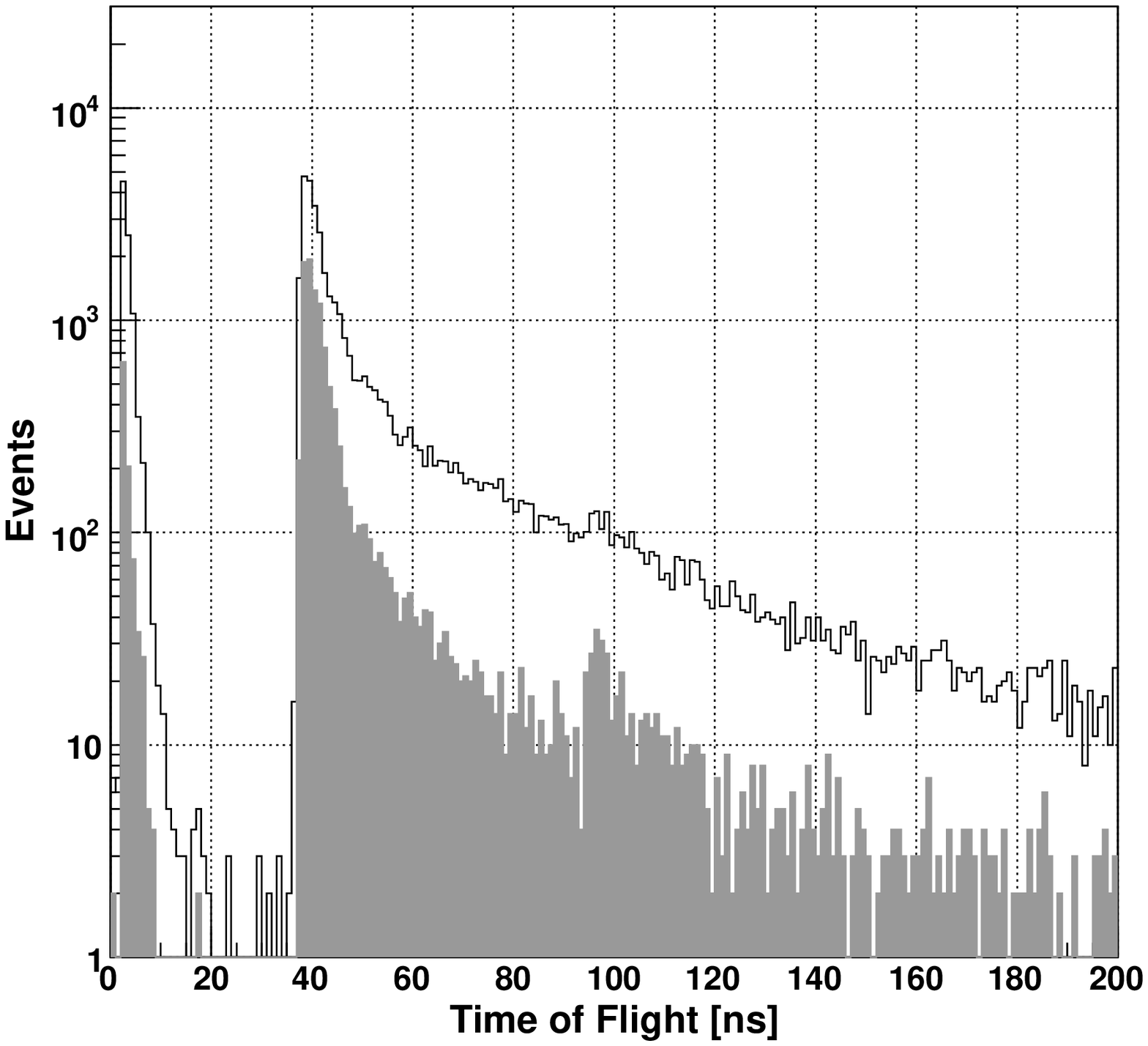} \\
      \mbox{(a)} & \mbox{(b)}
    \end{array}$
    \caption{Simulated time of flight distributions between NaI(Tl) and BC501A
      detector for scattering angles associated with: (a)~10~keVnr; and
      (b)~100~keVnr energy depositions. The unshaded histogram includes all
      events, while the shaded one only consists of single scattered events in
      the crystal. The peaks at 0~ns are from gamma-rays produced by inelastic
      scattering in the crystal. A decrease in events is seen at higher times.
      The delayed events are due to the scattering of neutrons off the wax
      shielding.}
    \label{tof_sim}
  \end{center}
\end{figure}

In Figure~\ref{tof_plots}, prior to the cut on BC501A pulse shapes outlined
previously, two peaks are visible at approximately 0~and 40~ns corresponding
to gamma-rays and neutrons, respectively. The time of flight differs from that
expected due to a time delay in the cables and the possibility of a neutron
to interact anywhere along the 8~cm depth of liquid scintillator that it
traverses. From Figure~\ref{qf_setup}, $s = 80$~cm is the distance from the
centre of the crystal to the face of the secondary detector. Substituting a
value of $s = 88$~cm into Eq.~(\ref{tof_eq}) results in an upper limit of
42~ns. This is confirmed from the simulated time of flight distributions for
scattering angles associated with 10~and 100~keVnr energy depositions in
Figure~\ref{tof_sim}. A sharp decline is witnessed in the number of events
that contribute to the simulated neutron peaks after 42~ns.

Due to the large background from gamma-rays, it is difficult to fit a Gaussian
function to the measured neutron peak in Figure~\ref{tof_plots}a. However,
after cutting on neutron events in the BC501A detector, as shown in
Figure~\ref{tof_plots}b, the neutron peak becomes clearly visible.

\subsection{Event selection by pulse shape discrimination in NaI(Tl)}

A variety of pulse shape discrimination techniques can be employed to
discriminate low energy nuclear and electron recoils in inorganic crystal
scintillators. Discrimination using mean time, neural networks and log
likelihood have been investigated in CsI(Tl) crystals~\cite{Wu04}. No
significant difference in efficiencies between the techniques was observed at
energy scales relevant to dark matter searches.

The scintillation mechanism of CsI(Tl) is similar to that of NaI(Tl), so there
is no reason to believe that the same result would not be true here. Therefore,
mean time is used for nuclear-electron recoil discrimination as it is the
easiest to implement with digitised waveforms. The reduction code calculates
the mean time $\langle t\rangle$ for each event with:

\begin{equation}
\langle t\rangle = \sum^{s_{2}}_{i=s_{1}}{\frac{A_{i}t_{i}}{A_{i}}}
\end{equation}
where $A_{i}$ is the amplitude of the digitised pulse and $t_{i}$ is the time
relative to the start of the pulse at sampling point $i$.

\begin{figure}
  \begin{center}
    $\begin{array}{c@{\hspace{5mm}}c}
      \includegraphics[scale=0.35]{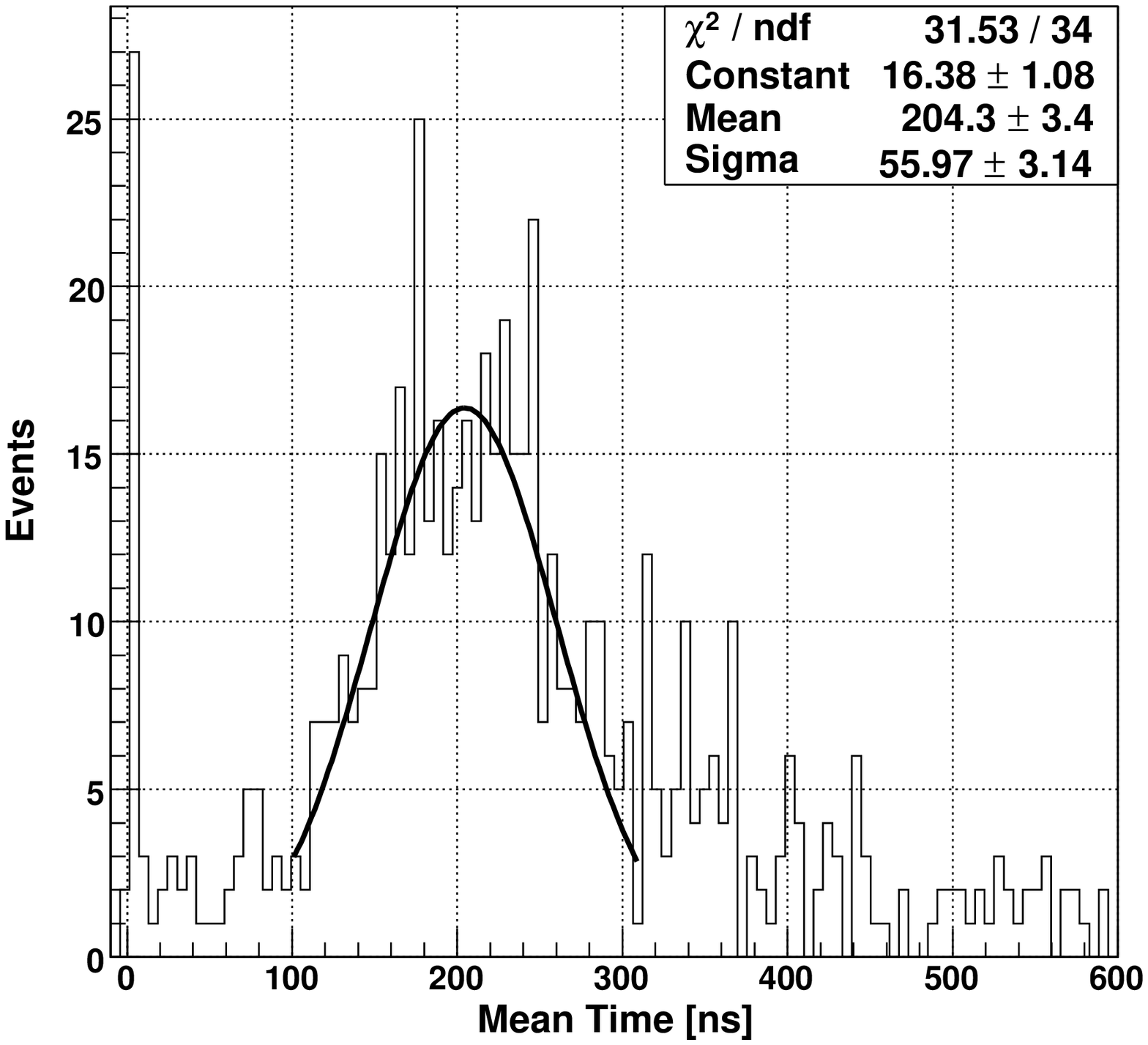} &
      \includegraphics[scale=0.35]{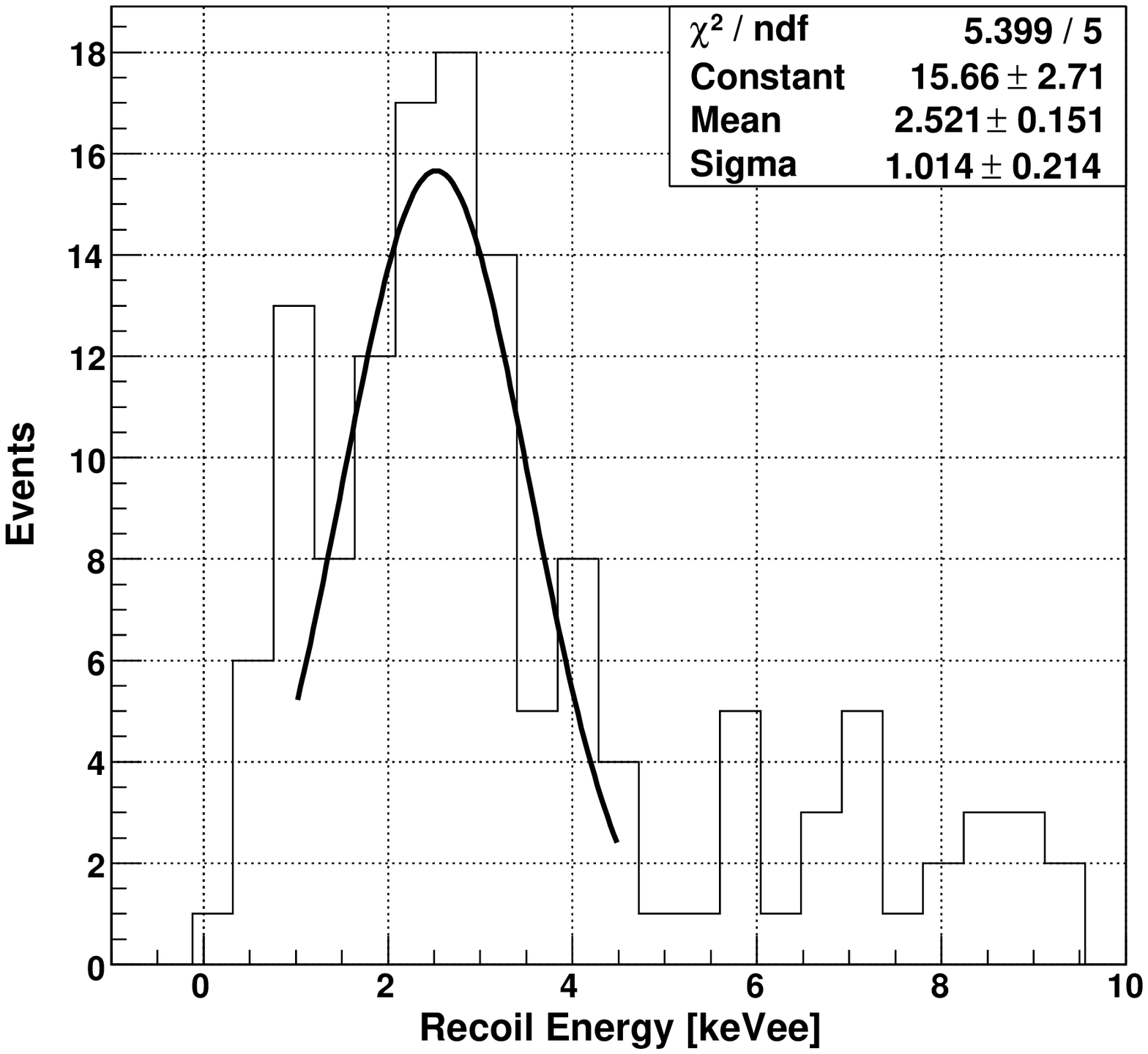} \\
      \mbox{(a)} & \mbox{(b)}
    \end{array}$
    \caption{(a)~Mean time of pulses from 10~keVnr Na recoils. (b)~Recoil
      energy in electron equivalent scale after events that lie more than half
      a standard deviation from the mean in (a) are excluded. The result in (b)
      indicates the quenching factor for 10~keVnr Na recoils in NaI(Tl) is
      25.21\%.}
    \label{10keVnr_res}
  \end{center}
\end{figure}

The vast majority of gamma events have been rejected by performing the lower
level cuts on PSD in BC501A and time of flight, as outlined above. 
Additional improvement is achieved by plotting mean time distributions of
events and removing those which lie more than a half of a standard deviation
from the peak position of the Gaussian fit (see Figure~\ref{10keVnr_res}a).
This is illustrated by the clear peak present in the resultant electron
equivalent energy distribution for 10~keVnr Na recoils in
Figure~\ref{10keVnr_res}b.

To assess the discrimination power of NaI(Tl), mean time distributions from
gamma-rays are compared with those from neutrons. The 511~keV gamma-ray from a
\isotope[22]{Na} source has a number of properties that make it attractive for
such a measurement. Neutrons will interact throughout the bulk of the crystal,
and gamma-rays of this energy have a typical penetration depth of 29~mm.
Additionally, the interaction cross-section in NaI(Tl) is dominated by Compton
scattering at this energy, meaning that a significant number of gamma-rays will
deposit a fraction of their energy in the crystal before escaping. Finally,
their back-to-back emission is exploited by placing the source between the
crystal and BC501A detector, and operating them in coincidence using the
electronics shown in Figure~\ref{nbeam_elec}. The width of NIM pulses from the
discriminator is reduced to 10~ns, which is the minimum setting, as coincident
gammas will arrive at each detector at the same time. This enables the
discriminator threshold to be decreased to 2~mV without noisy events polluting
the data.

\begin{figure}
  \begin{center}
    \includegraphics[scale=0.7]{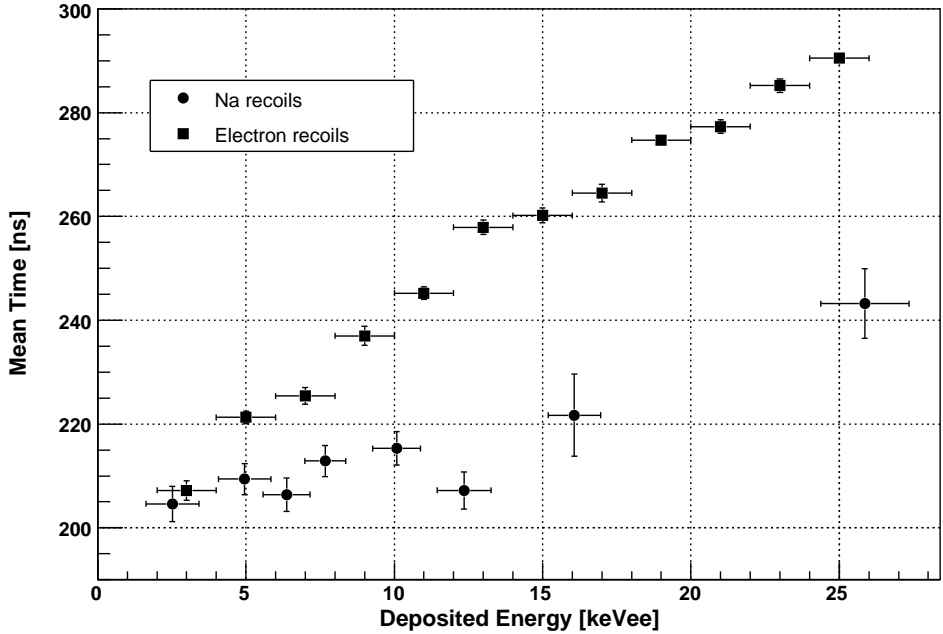}
    \caption{Mean time as a function of deposited energy for sodium (Na) and
      electron recoils. Measurements of Na recoils are performed with the
      neutron beam. Compton electrons are induced by gamma-rays from a
      \isotope[22]{Na} source.}
    \label{tau_vs_nrg}
  \end{center}
\end{figure}

Mean time values for electron recoils are evaluated in 2~keVee wide energy
bins, each containing approximately 6~000~events. Values for nuclear recoils
come from the data taken with the neutron beam at each scattering angle (see
Table~\ref{QF_tab}). The results, from Figure~\ref{tau_vs_nrg}, indicate that
mean time values for sodium recoils stay roughly constant with energy (only
small increase is seen above 15~keV), compared with those from Compton
scatters (electron recoils) that increase significantly, in agreement
with~\cite{Tovey98}~\cite{Gerbier99}. Additionally, it becomes difficult to
distinguish electron and nuclear recoils at energies below 4~keVee, hampering
the sensitivity of NaI(Tl) dark matter detectors.

\section{Results}

\begin{figure}
  \begin{center}
    \includegraphics[scale=0.7]{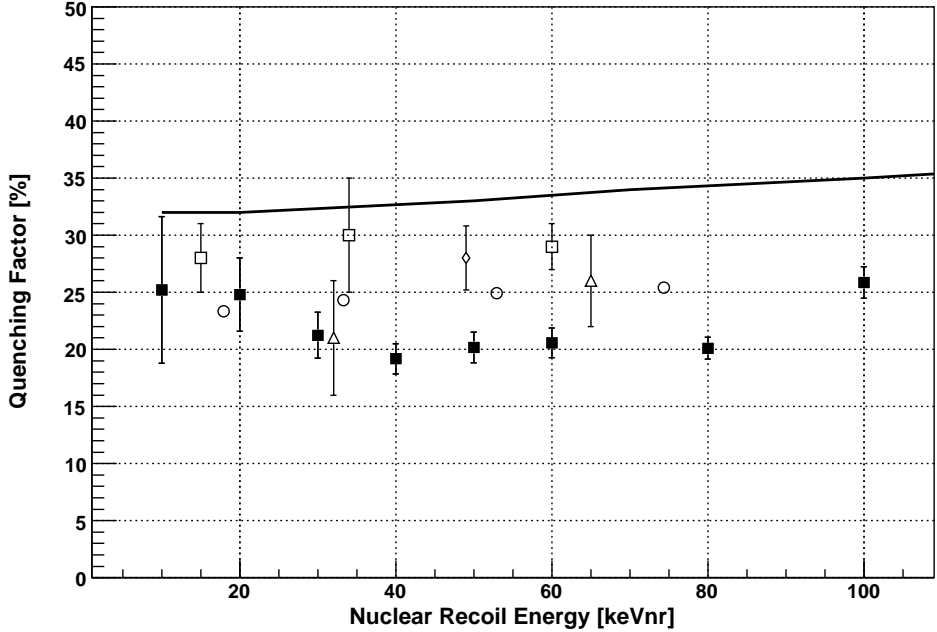}
    \caption{Quenching factor of Na recoils in NaI(Tl). Experimental results
      from this work (filled black squares), Spooner et al.~\cite{Spooner94}
      (open squares), Tovey et al.~\cite{Tovey98} (open triangles), Gerbier et
      al.~\cite{Gerbier99} (open circles) and Simon et al.~\cite{Simon03} (open
      diamond) are shown. Additionally, the preliminary theoretical estimation
      of the quenching factor from Hitachi~\cite{Hitachi06} is represented by
      the solid black line.}
    \label{qf_res}
  \end{center}
\end{figure}

Distributions similar to that shown in Figure \ref{10keVnr_res}b are
constructed for different energy bins in keVnr scale. The areas around the
peaks are fitted to the Gaussian function, the peak positions of the fits
being associated with the measured (or electron equivalent) energy. The
quenching factors for each energy bin are determined as the ratio of electron
equivalent to recoil energy. Values of the measured energies and resultant
quenching factors at each scattering angle are given in Table~\ref{QF_tab}.
The quenching factor of sodium recoils in NaI(Tl) varies between 19\% and 26\%
in the range 10~to 100~keVnr, in agreement with previous experimental
results~\cite{Spooner94}~\cite{Tovey98}~\cite{Gerbier99}~\cite{Simon03}, as
shown in Figure~\ref{qf_res}. A scintillation efficiency of $25.2 \pm 6.4$\%
has been determined for 10~keVnr Na recoils.

\begin{table}
\begin{center}
\begin{tabular}{cccc}
\hline
Scattering angle & Recoil energy & Measured energy & Quenching factor \\
$\theta$ [$^{\circ}$] & $E_{R}$ [keVnr] & $E_{\mathrm{vis}}$ [keVee] &
$Q$ [\%] \\
\hline\hline
18.4 & 10 & $2.52 \pm 0.90$ & $25.2 \pm 6.4$ \\
26.1 & 20 & $4.96 \pm 0.88$ & $24.8 \pm 3.2$ \\
32.1 & 30 & $6.37 \pm 0.79$ & $21.2 \pm 2.0$ \\
37.3 & 40 & $7.67 \pm 0.69$ & $19.2 \pm 1.3$ \\
41.9 & 50 & $10.08 \pm 0.81$ & $20.2 \pm 1.3$ \\
46.1 & 60 & $12.35 \pm 0.91$ & $20.6 \pm 1.3$ \\
53.8 & 80 & $16.07 \pm 0.88$ & $20.1 \pm 1.0$ \\
60.7 & 100 & $25.86 \pm 1.49$ & $25.9 \pm 1.4$ \\
\hline
\end{tabular}
\caption{Quenching factors of Na nuclear recoils relative to those of
gamma-rays of the same energy. The average scattering angle is given by
$\theta$ (column~1), the average recoil of the Na nucleus is given by $E_{R}$
(column~2) and the measured energy is given by $E_{\mathrm{vis}}$ (column~3).
The fractional contribution from statistical errors remains constant at
$\approx 0.05$ over the full energy range. The systematic error is dominated by
uncertainties in the determination of the scattering angle. A marked increase
in the contribution from the systematic error is seen at low $\theta$, where
$E_R < 20$~keVnr. At higher recoil energies, a reduction in the relative
contribution of the systematic error is seen, as it decreases by just under an
order of magnitude over the full energy range. Systematic and statistical
errors are added quadratically to obtain the uncertainty on the quenching
factor.}
\label{QF_tab}
\end{center}
\end{table}

Systematic errors in the measurement of the scattering angle start to dominate
at energies less than 20~keVnr. Although it may be possible to take a
measurement at 5~keVnr, especially as the light yield seems to increase at
lower nuclear recoil energies, the magnitude of the systematic error at such a
scattering angle would be too large to obtain a sensible result. Therefore, the
limiting factor in this experiment is not the light yield, but the error
associated with the scattering angle.

There are a number of features in Figure~\ref{qf_res}, including a dip in the
quenching factor around a nuclear recoil energy of 40~keVnr, and subsequent
rise at lower energies. This is the first time a dip has been observed. The
measurement performed here is the most comprehensive study of the quenching
factor of sodium recoils in NaI(Tl) to date in the low energy regime. It is
therefore possible that such a feature could have been hidden from other
experiments, as there are fewer data points available to witness this pattern.
A similar trend has been seen in liquid xenon at energies below
10~keVnr~\cite{Chepel06}, indicating that there is some underlying process
responsible for these observations.

Quenching factors for silicon recoils in Si, argon in Ar, germanium in Ge and
xenon in Xe have been derived from SRIM by~\cite{Mangiarotti07}, and compared
with predictions from Lindhard theory and experimental data where available.
Their results indicate that the nuclear stopping powers predicted by Lindhard
theory and calculated by SRIM differ by 15\% at most, although bigger
discrepancies are present for the electronic stopping power. When compared
with experimental data, the original Lindhard theory is closest to giving an
accurate prediction for these media.

Neither Lindhard theory nor the results from SRIM reproduce the shape of the
experimental results for Na recoils in NaI(Tl). Unlike the prediction from
Hitachi~\cite{Hitachi06}, which provides a better resemblance to the pattern
seen, they do not consider the effect of electronic quenching due to high LET
of ions. However, the appearance of the dip remains unexplained.

\section{Conclusion}
Quenching factor measurements have been performed for sodium recoils in a 5~cm
diameter, cylindrical NaI(Tl) crystal. The results show an average quenching
factor of 22.1\% at energies less than 50~keVnr, in agreement with other
measurements. Results from simulations confirm that the contribution from
multiple scattering events provides a featureless background, and can be
neglected. The results do not reproduce the shape of the predicted curves from
Lindhard theory, and SRIM and TRIM. However, the predicted quenching factor
from Hitachi~\cite{Hitachi06},  which takes electronic quenching into account,
compares favourably with the experimental results. The presence of a dip in
the quenching factor at around 40~keVnr is observed.

\acknowledgments
The authors would like to thank Prof. Akira Hitachi for valuable discussions
of the results, and the provision of a preliminary theoretical prediction of
the quenching factor. HC would also like to thank STFC (formally PPARC) for
the support of a PhD studentship. This work has partly been supported by the
ILIAS integrating activity (Contract No. RII3-CT-2004-506222) as part of the
EU FP6 programme in Astroparticle Physics.

\end{document}